\documentclass[twocolumn,10pt,twoside]{IEEEtran}

\usepackage{amsmath,amsthm,graphicx,bm}
\usepackage{epsfig,amsfonts}
\usepackage{color}
\usepackage[dvipsnames]{xcolor}
\usepackage[nolist]{acronym}
\usepackage{psfrag}
\usepackage{perso}
\usepackage{xspace}
\usepackage{bbding}
\usepackage{anyfontsize}

\usepackage{pgfplots}
 \pgfplotsset{compat=newest}
    \pgfplotsset{plot coordinates/math parser=false}
    \pgfplotsset{
    label style={anchor=near ticklabel},
    xlabel style={yshift=0.0em},
    ylabel style={yshift=-0.3em},
    tick label style={font=\footnotesize },
    label style={font=\footnotesize},
    legend style={font=\footnotesize},
    title style={font=\fontsize{7}}}

\usepackage{mathtools}
\mathtoolsset{showonlyrefs}

\usepackage{placeins}

\usepackage{multirow}
\usepackage{etoolbox}

\usepackage{flushend}
\usepackage{placeins} 

\usepackage{algpseudocode}
\algrenewcommand\algorithmicrequire{\textbf{Input:}}
\algrenewcommand\algorithmicensure{\textbf{Output:}}

\usepackage{subfig}
\usepackage[normalem]{ulem}

\usepackage{amsthm}
\newtheorem{example}{Example}
\usepackage{booktabs}
\usepackage{booktabs}
\usepackage{ulem}

\usepackage{slashbox}

\usepackage[linesnumbered,ruled,vlined]{algorithm2e}  

\usepackage{algorithmicx}
\usepackage{algpseudocode}
\SetKwComment{tcp}{\small/*}{*/}
\SetKwProg{Fn}{Function}{:}{}

\newcommand{\overbar}[1]{\mkern 2.5mu\overline{\mkern-2.5mu#1\mkern-2.5mu}\mkern 2.5mu}

\renewcommand{\S}{\mathcal{S}}
\newcommand{\SA}{\S_A}
\newcommand{\SB}{\S_B}

\renewcommand{\P}{\mathcal{P}}

\newcommand{\D}{\mathcal{D}}

\renewcommand{\d}{\delta}

\newcommand{\elsize}{\ell}

\newcommand{\Z}{\mathcal{Z}}
\newcommand{\Zp}{\mathcal{Z}_\mathrm{p}}

\newcommand{\flag}{\mathtt{flag}}
\newcommand{\flagskip}{\mathtt{skip}}

\newcommand{\redX}{\textcolor{BrickRed}{\XSolidBrush}}
\newcommand{\greencheck}{\textcolor{ForestGreen}{\CheckmarkBold}}

\newcommand{\bardelta}{{\bar{m}}}
\newcommand{\kfact}{\gamma}

\newcommand{\req}{\mathtt{request}}

\newcommand{\pmax}{p_{\mathrm{max}}}
\newcommand{\qmax}{q_{\mathrm{max}}}
\newcommand{\qmin}{q_{\mathrm{min}}}

\newcommand{\depth}{d}
\newcommand{\Depth}{D}

\newcommand{\cost}{\mathtt{cost}}

\renewcommand{\Pr}{\mathrm{P}}

\newcommand{\Pf}{\mathrm{P}_{\mathrm{F}}}

\makeatletter
\newcommand{\algcolor}[2]{%
  \hskip-\ALG@thistlm\colorbox{#1}{\parbox{\dimexpr\linewidth-2\fboxsep}{\hskip\ALG@thistlm\relax #2}}%
}

\makeatother

\newcommand{\npc}{N}
\newcommand{\expnpc}{\overbar{\npc}}

\newcommand{\npce}{T}
\newcommand{\expnpce}{\overbar{\npce}}

\newcommand{\nrec}{U}
\newcommand{\expnrec}{\overbar{\nrec}}

\newcommand{\mypgf}[1]{G_{#1}}

\newcommand{\rec}{{\emph{recovery}} }

\newtheorem{theorem}{Theorem}
\newtheorem{corollary}{Corollary}
\newtheorem{lemma}{Lemma}

\newtheorem{definition}{Definition}

\newcommand{\epsrskip}{h}

\begin{document}

\usetikzlibrary{patterns}

\usetikzlibrary{arrows.meta,arrows}

\usetikzlibrary{shapes.geometric}
\tikzset{database/.style={cylinder,aspect=0.5,draw,rotate=90,path picture={
\draw (path picture bounding box.160) to[out=180,in=180] (path picture bounding
box.20);
\draw (path picture bounding box.200) to[out=180,in=180] (path picture bounding
box.340);
}}}

\tikzstyle{block} = [draw, fill=white!20, rectangle, rounded corners,  minimum height=3em, minimum width=7em, text width=6em, text centered]

\usetikzlibrary{positioning}

\usetikzlibrary{calc}

\tikzset{
  rectangle with ellipse/.style={
    draw, rectangle, minimum width=##4, minimum height=##2, 
    append after command={
      \pgfextra{
        \draw[thick] (current bounding box.center) ellipse (##1/3 and ##2/4);
        \node at (current bounding box.center) {$\mathcal{S}$};
        \node at (current bounding box.south east) [xshift=-0.5cm, yshift=0.5cm] {$\mathcal{U}$};
      }
    }
  }
}

\newcommand{\drawVenn}[2]{    
        \draw (#1) rectangle ++(4, 3);

            \clip (#1) rectangle ++(4, 3);
            \draw[fill=none] (#1) ++(1.5, 1.5) circle (1.2) node[below left] {A};
            \draw[fill=none] (#1) ++(2.5, 1.5) circle (1.2) node[below right] {B};
            
}

\tikzset{
  thick vertical lines/.style={
    pattern=custom north west lines,
    pattern color=black,
    pattern line width=1.5pt  
  }
}

\usetikzlibrary{patterns.meta}

\SetKwInput{Input}{Input}
\SetKwInput{Output}{Output}
\SetKwBlock{Spawn}{\textbf{spawn}}{\textbf{end spawn}} 
\SetKw{WaitAllChildren}{waitAllChildren}               
\SetKwFor{ParFor}{parallel for}{do}{end parallel for} 
\newcommand{\atomic}[1]{\textbf{atomic}\; #1}   
 
	\begin{acronym}
\acro{CPI}{characteristic polynomial interpolation}
\acro{PSR}{partitioned set reconciliation}
\acro{BSR}{basic set reconciliation}
\acro{EPSR}{enhanced partitioned set reconciliation}
\acro{SR}{set representation }
\acro{r.v.}{random variable}
\acro{PGF}{probability generation function}

\acro{rhs}{right-hand side}

\acro{w.r.t.}{with respect to}

\end{acronym}

\title{Tree algorithms for set reconciliation}

\author{
	Francisco L\'azaro, \IEEEmembership{Senior Member, IEEE},\thanks{Francisco L\'azaro is with the Institute of Communications and Navigation of the German Aerospace Center (DLR), Muenchner Strasse 20, 82234 Wessling, Germany. Email: \texttt{Francisco.LazaroBlasco@dlr.de}.} 
         \v Cedomir Stefanovi\' c, \IEEEmembership{Senior Member, IEEE}	
	\thanks{\v Cedomir Stefanovi\' c is with the Department of Electronic Systems of Aalborg University, 2450 Copenhagen, Denmark.
		E-mail: \texttt{cs@es.aau.dk}.}    
	\thanks{Corresponding Address:
		Francisco L\'azaro, KN-SAN, DLR, Muenchner Strasse 20, 82234 Wessling, Germany.  E-mail: \texttt{Francisco.LazaroBlasco@dlr.de}.}
    
	\thanks{This work has been submitted to the IEEE for possible publication. Copyright may be transferred without notice, after which this version may no longer be accessible.}
}

\maketitle

\thispagestyle{empty} \pagestyle{empty}

	\begin{abstract}
        In this work, a set reconciliation setting is considered in which two parties have similar sets that they would like to reconcile.
        In particular, we focus on a divide-and-conquer strategy known as \acf{PSR}, in which the sets to be reconciled are successively partitioned until they contain a number of differences below some predetermined value.
        Borrowing techniques from tree-algorithms for random-access protocols,         
        we present and analyze a novel set reconciliation scheme that we term \acf{EPSR}.
        This approach improves the efficiency in terms of overhead, i.e., it yields a lower communication cost, while keeping the same time and communication round complexity as \ac{PSR}.
        Additionally, we simulate the performance of the proposed algorithm in an event-driven simulator. 
        Our findings indicate that this novel protocol nearly halves the communication cost of \ac{PSR} while maintaining the same time complexity.                        
\end{abstract}

\maketitle
	\thispagestyle{plain}
    \pagestyle{plain}
		\setlength{\footskip}{12.3pt}

\section{Introduction}\label{sec:Intro}

In distributed systems, there are often scenarios where multiple nodes hold slightly different versions of similar datasets, such as a database, that need to be synchronized.
This setup is typically framed as a set reconciliation problem \cite{Karpovsky2003}, where two nodes $A$ and $B$ maintain each a set of elements (denoted $\SA$ and $\SB$) and wish to compute the difference between their sets while exchanging as little data as possible.

The applications of set reconciliation algorithms are not limited to distributed databases. For example, they can also be used to build remote file synchronization protocols \cite{agarwal2006bandwidth}, or to improve the efficiency of gossip protocols \cite{MitzenmacherP18}. Set reconciliation algorithms have also been applied to solve data-synchronization problems in cryptocurrency networks \cite{Ozisik2019,bovskov2022gensync,yang2024practical}. 
Additionally, a recent work has proposed using set reconciliation algorithms to improve the performance of end-to-end transport protocols by providing a means of detecting lost packets \cite{yuan2024}.

The set reconciliation problem has been extensively explored in the literature, with several methods achieving near-optimal communication costs, though often with significant computational complexity \cite{Minsky2003, dodis2004fuzzy}.

Among the myriad of algorithms for set reconciliation presented in the literature, this study concentrates on refining a divide-and-conquer approach known as \acf{PSR} \cite{Minsky2002}. This method offers the significant benefit of linear complexity in relation to the number of differences in the sets, though it incurs a higher communication cost compared to the optimal solutions and necessitates multiple rounds of communication.
This paper introduces an innovative algorithm called \ac{EPSR}, which improves the communication efficiency of \ac{PSR} by incorporating strategies from tree-based algorithms used in random-access protocols. The proposed algorithm is especially beneficial for synchronizing sets with many differences in low-latency networks, achieving a communication cost close to optimal solutions while maintaining a significantly lower time complexity.

The structure of the paper is as follows. Section~\ref{sec:sys_model} provides an overview of the system model and its relation to existing work. Section~\ref{sec:prelim} introduces the fundamental set reconciliation primitive employed in this study and reviews \ac{PSR}, the baseline algorithm targeted for enhancement. Section~\ref{sec:enhanced} presents the proposed set reconciliation algorithm, referred to as \ac{EPSR}. Section~\ref{sec:analysis} offers an exact recursive analysis of the communication cost and time complexity of both \ac{PSR} and \ac{EPSR}.
Section~\ref{sec:num_res} presents numerical results used to validate our analysis, as well as the simulated performance of the algorithms in a controlled network environment. Finally, conclusions are drawn in Section~\ref{sec:conclusions}.

    \section{System Model and Related Work}\label{sec:sys_model}

\subsection{System Model}

We consider a setup with two nodes, $A$ and $B$, each of them having a set $\SA$ and $\SB$, and let the elements of $\SA$ and $\SB$ be $\elsize$ bit vectors, i.e., $\SA, \SB\subset \{0,1\}^\elsize$. 
Our goal is to determine the set difference $\D$, i.e., the set of elements present in the union of $\SA$ and $\SB$, but not in their intersection: $\D = \SA \ominus \SB = (\SA \cup \SB) \setminus (\SA \cap \SB)$.
The cardinality of the set difference will be denoted by $\d=|\D|$.

To reconcile their sets, nodes A and B follow a set reconciliation protocol, that is, a distributed algorithm that involves the exchange of information and performing calculations to determine $\D$. 
More specifically, we assume that node $B$ acts as \emph{transmitter} and sends information to node $A$, possibly over multiple communication rounds. Node $A$ acts as \emph{receiver} and processes the information received from $B$ to derive the set difference $\D$. Additionally, node $A$ might send short feedback messages to node $B$ requesting additional information.

To measure the performance of a set reconciliation algorithm, we shall focus on its communication complexity (measured in bits), its computational (time) complexity, and also on the number of communication rounds\footnote{Within a single communication round, node A simultaneously sends one or more request messages to node B, and node B subsequently sends the corresponding reply messages back to node A.} required to synchronize the sets.
Regarding communication complexity, we will take into account only the messages sent from $B$ (transmitter) to $A$ (receiver), neglecting the feedback messages sent from $A$ to $B$ since they are very short.

\subsection{Related Work}

The fundamental limits for one-way set reconciliation, when one single communication message is allowed, were studied in \cite{Karpovsky2003}. In this work, it was found that, when an upper bound $t$ to the set difference $\d$ is known, exact set reconciliation is possible with communication complexity $\sim t \elsize$.
In \cite{Minsky2003}, a scheme based on \acf{CPI} was proposed that achieves nearly optimal communication complexity. In particular, when an upper bound $t$ to the size of the set difference $\d$ is known, exact set reconciliation is possible with communication complexity $\sim t (\elsize+1)$ and time complexity $\mathcal{O}(t^3 )$. 
When an upper bound to $\d$ is not known, \cite{Minsky2003} proposes an approximate\footnote{Reconciliation may fail, so that the parties wrongly believe that their sets are reconciled, but they actually are not. However, the probability of failure can be made arbitrarily low, at the expense of an increase in communication cost.} set reconciliation scheme with communication complexity $\sim \d (\elsize+1)$ and time complexity $\mathcal{O}(\d^4)$.
In \cite{dodis2004fuzzy}, a scheme known as PinSketch was introduced, which improves on the performance of \ac{CPI}. PinSketch is based on BCH codes and it has communication complexity  $\sim t \ell$ and time complexity  $\mathcal{O}(t^2)$. PinSketch, however, requires knowledge of an upper bound $t$ to $\d$.

In practice, \ac{CPI} and PinSketch are a suitable solution to reconcile sets as long as the number of differences $\d$ is moderate (at most in the order of several hundreds). When the number of differences $\d$ is large, one typically resorts to other, faster, approximate set reconciliation algorithms based on probabilistic data structures. For example, schemes based on Bloom filters, counting Bloom filters, and Cuckoo filters were proposed in \cite{Skjegstad2011},  \cite{Guo2013} and \cite{Luo2019}, respectively, whereas schemes based on invertible Bloom look-up tables were proposed in \cite{eppstein:2011,lazaro21:iblt, lazaro2022setrec, yang2024practical}.
Alternatively, as proposed in \cite{Minsky2002}, it is also possible to rely on  a divide-and-conquer approach that builds on a primitive set reconciliation algorithm (e.g., \ac{CPI}). 
This approach is termed \acf{PSR} and consists of recursively partitioning the sets $\SA$ and $\SB$ into smaller disjoint subsets, which can then be reconciled with lower complexity since they contain fewer differences. 
This method features linear communication and time complexity, but its communication complexity is, at best, roughly three times greater than that of \ac{CPI}, and thus far from optimal.

\subsection{Contribution}
Inspired by a family of random-access algorithms known as \emph{tree-algorithms} \cite{capetanakis1979tree,yu2007high,stefanovic2023tree,Flajolet,SGDK2020},
we introduce a novel set reconciliation scheme, 
termed \acf{EPSR}.
As the name suggests, \ac{EPSR} builds upon and enhances \ac{PSR}.
The working principle of \ac{PSR} is the following: when the number of differences between the two sets is too large (above the predefined bound $\bardelta$), \ac{PSR} partitions the current set recursively into smaller subsets (child partitions) until the number of differences in a subset falls below $\bar m$. At each split, node~A requests from node~B a so called \acf{SR} data structure for \emph{every} child partition and attempts recovery on all children in parallel.
\ac{EPSR} exploits the fact that the \ac{SR} data structure of a parent partition can be \emph{reused} for one child without any additional transmission. Concretely, when a partition has to be split, node~A (i) requests from node~B one single \ac{SR} data structure for a first child partition, (ii) uses the parent \ac{SR} to obtain an additional \ac{SR} data structure for a second child partition without having to transmit it over the network and (iii) processes these two child partitions in parallel. 
This improvement reduces the number of \ac{SR} data structure transmissions, and thus the communication cost.

Additionally, we propose a novel recursive analysis of the average communication cost and time complexity of \ac{PSR}. In contrast to the analysis provided in \cite{Minsky2002}, our approach is exact, recursive in nature, and accommodates arbitrary partition sizes.
Subsequently, this analysis is also extended to \ac{EPSR}. 

    \section{Preliminaries}
\label{sec:prelim}

\subsection{Set Representation Data Structure}\label{sec:data_structure}

The algorithms considered in this paper rely on a so-called \acf{SR} data structure $\Z$, which can be used to perform the following operations on a set $\S =\{s_1, s_2, \dots, s_u\}$:
\begin{itemize}
    \item $\Z$.\emph{init}: creates an \emph{empty} \ac{SR} data structure.
    \item $\Z$.\emph{insert\_el$(s)$}: embeds    
    a set element $s$ in $\Z$.
    \item $\Z$.\emph{insert\_set$(\S)$}: embeds all elements $s$ of set $\S$ in $\Z$ relying on the operation \emph{insert\_el$(s)$}.    
    \item $(\text{\emph{flag}}, \mathcal{L}) =\Z$.\emph{recovery}(): attempts to recover all set elements that are represented in $\Z$. 
    This function provides two outputs. The first output, \emph{flag}, is a Boolean variable that is assigned the value \emph{true} if the function is successful. In contrast, if the function fails, the value of \emph{flag} is set to \emph{false}.
    The second argument $\mathcal{L}$  is the set elements that have been recovered.
    If recovery is successful, we have $\mathcal{L}= \Z$, otherwise we have $\mathcal{L} \subset \Z$.
    \item $\Z_{A \ominus B}=$\emph{subtract}$(\Z_A, \Z_B)$: subtracts the data structure $\Z_B$ from $\Z_A$, resulting in a third data structure $\Z_{A \ominus B}$. If data structures $\mathcal{Z}_A$ and $\Z_B$ represent respectively sets $\SA$ and $\SB$, $\Z_{A \ominus B}$ is a data structure representing their (symmetric) set difference $\D= \SA \ominus \SB$. 
\end{itemize}
In short, $\Z$ is a data structure that can be seen as a sort of check-sum or summary that allows us to compactly represent a set and on which we can carry out some set operations.

In this work, we will assume that the data structure $\Z$ is based on \acf{CPI}~\cite{Minsky2003}.\footnote{It is also possible to rely on invertible Bloom lookup tables \cite{goodrich:2011, lazaro21:iblt, bar-lev:2023}, or BCH codes \cite{dodis2004fuzzy} to create $\Z$.} 
The \ac{CPI} set representation data structure can be configured based on two parameters $\bardelta$ and $\kfact$. 
The first parameter, $\bardelta$, controls the number of elements that can (unequivocally) be represented in the data structure. More concretely, if $\mathcal{Z}$ represents a set $\S$,  we have that recovery will always be successful whenever $|\S| \leq \bardelta$. 
In contrast, when $|\S| > \bardelta$ recovery fails. 
Here, two types of failures can be distinguished: true failures and false successes. In a true failure, recovery returns \emph{flag=false} (as it should), whereas in a false success we have \emph{flag=true} but the returned set $\mathcal{L}$ does not correspond to the set $\S$ represented in the \ac{SR} data structure.
The second parameter, $\kfact$, is used to control the false success probability, $\Pf$. Specifically, $\Pf$ is given by \cite{Minsky2002}
\[
\Pf=\left( \frac{ |\S| } { 2^\ell}\right)^{\kfact}.
\]
Note that, typically, $2^\ell >> |\SA| + |\SB|$. 
Thus, $\Pf$ can be made small enough by setting $\kfact$ to a sufficiently large value\footnote{Usually, setting $\kfact$ to 1 or 2 is adequate.}. 

The communication cost associated with the transmission of the data structure $\mathcal{Z}$ (i.e., its size in bits) depends on the parameters $\bardelta$ and $\kfact$, as well as the number of bits $\elsize$ needed to represent a set element. More concretely, we have that the size of  $\mathcal{Z}$ in bits is given by  \cite{Minsky2002}
\begin{equation}
   \cost_\Z(\bardelta,\kfact) = (\bardelta+\kfact+1) (\ell +1) -1
   \label{eq:cost}
\end{equation}
The time (or computational) complexity of the different operations on the data structure $\mathcal{Z}$ is shown in Table~\ref{tab:complexity}.

\begin{table}[t]
    \centering
    \begin{tabular}{|c|c|}\hline
         operation & time complexity\\\hline
         \emph{init} & $\mathcal{O}(\ell (\bardelta + \kfact))$ \\ \hline
         \emph{insert\_el} & $\mathcal{O}(\ell (\bardelta + \kfact))$ \\ \hline         
         \emph{recovery} & $\mathcal{O}(\ell \bardelta^3 + \ell \bardelta \kfact)$ \\ \hline
         \emph{subtract} & $\mathcal{O}(\ell (\bardelta + \kfact))$ \\ \hline
    \end{tabular}
    \caption{Time or computational complexity of the different operations on the data structure $\mathcal{Z}$.}
    \label{tab:complexity}
\end{table}

    \subsection{Partitioned Set Reconciliation}

\Acf{PSR} \cite{Minsky2002} solves the set reconciliation algorithm using a divide-and-conquer approach and relying on \ac{SR} data structures. 
Initially, an attempt to reconcile the sets $\SA$ and $\SB$ is carried out relying on the $\rec$ operation.
If reconciliation succeeds at this level, the sets are reconciled in their entirety. However, if reconciliation fails, the sets  $\SA$ and  $\SB$ are partitioned into smaller subsets. These subsets are then reconciled independently and in parallel, as their reconciliation processes do not interdepend.

Formally, a $c$-ary partitioning of a set $\P$  is a sequence $\{\P_1, \P_2, \dots \P_c \}$  of mutually disjoint subsets of $\P$, $\P_i \cap \P_j=\emptyset$ for $i\neq j$, whose union coincides with $\P$, $\cup_{i=1}^c \P_i = \P$.
Let $\{\P_1, \P_2, \dots \P_c \}$ be a partitioning of $\P$ and consider two sets $\SA$, $\SB \subset \P$. The symmetric difference between the sets, $\D = \SA \ominus \SB$ can be recovered by processing separately all partitions,
\[
\D = \SA \ominus \SB = \bigcup_{i=1}^c (\SA \cap \P_i) \ominus  (\SB \cap \P_i). 
\]
This process is repeated as many times as necessary until reconciliation succeeds, see Algorithm~\ref{alg:psr} for a detailed description.

The  \ac{PSR} protocol has three parameters: the \ac{SR} data structure parameters $\bardelta$ and $\kfact$ and additionally a \emph{partitioning schedule} that defines how the universal set $\P$ is recursively partitioned. 
In this paper, we will assume that a set is always partitioned into $c$ partitions, that need not necessarily be of equal size.

Algorithm~\ref{alg:psr} shows a pseudocode description of \ac{PSR} as a recursive algorithm. 
In this description, line 3 ($\Z_B \gets \req(\SB \cap \P)$ ) issues a request from node A to node B over the data network to obtain the \ac{SR} data structure associated with partition $\SB \cap \P$. Following this request, node B sends the corresponding \ac{SR} data structure back over the network, incurring the communication cost in Eq.~\eqref{eq:cost}.
The statement annotated as \emph{atomic} executes as a critical section with respect to the data structure that represents the set $\D$, i.e., while such a statement is executing, no other process may read or modify that data structure.

The communication cost of one call to \ac{PSR} corresponds to the size of the \ac{SR} data structure that needs to be transmitted, which is $ (\bardelta+\kfact+1) (\ell +1) -1$, see Eq.~\eqref{eq:cost}.
The computational time complexity of \ac{PSR} can be inferred from Table~\ref{tab:complexity}. 
Typically, when $\bardelta$ is large, the time complexity is dominated by the \rec operation, whose complexity is  $\mathcal{O}(\bardelta^3)$.
Hence, the total complexity and communication cost can be derived by tracking how many times \ac{PSR} invokes itself (see Section~\ref{sec:analysis}).

\begin{algorithm} 
\caption{$[\D] = \textsc{PSR}(\SA, \P)$}
\label{alg:psr}
\Input{

  $\SA$ - Host A’s set\;  
  $\P$  - partition to process
}
\Output{$\D = (\SA\cap\P)\ominus(\SB\cap\P)$}

\BlankLine
$\D \gets \emptyset$\;

$\Z_A.\mathsf{insert\_set}(\SA  \cap \P)$\;
$\Z_B \gets \req(\SB \cap \P)$ 

$\Z \gets \mathsf{subtract}(\Z_A,\Z_B)$\;

$[\flag,\D] \gets \Z.\mathsf{recovery}()$\;
\If{$\flag$}{\Return{$\D$}}

$\{\P_1,\dots,\P_c\} \gets \mathsf{Partition}(\P)$\;

\ParFor{$i \gets 1$ \KwTo $c$}{
    $[\D_i] \gets \textsc{PSR}(\SA,\P_i)$\;
    $\D \gets \D \cup \D_i$ \tcp*[r]{atomic}
} 

\Return{$\D$}\;
\end{algorithm}

\begin{example}[\Acf{PSR}]\label{example:PSR}
Consider the example shown in Fig.~\ref{fig:tree_no_sic_m_2} where \acs{PSR} is instantiated with $\bardelta=2$ and $c=2$.
The \#\textcolor{gray}{root} partition at level 0 has $9$ differences, causing $\rec$ to fail. Consequently, it is split into two child partitions: left (\#\textcolor{gray}{l}) and right (\#\textcolor{gray}{r}), which are at level 1.
Next, the two partitions at level 1, \#\textcolor{gray}{l} and \#\textcolor{gray}{r} are processed concurrently (i.e., independently). Since both partitions contain more than $\bardelta=2$ differences, $\rec$ fails for both of them, and the partitions are split again. 
Let us focus first on partition \#\textcolor{gray}{l}, which is split into two partitions, labeled \#\textcolor{gray}{ll} and \#\textcolor{gray}{lr}.
Partition \#\textcolor{gray}{ll} contains only 2 differences, so $\rec$ succeeds, and there is no need to do a further splitting.
However, partition \#\textcolor{gray}{lr} contains 3 differences, so $\rec$ fails and the partition is split into partitions \#\textcolor{gray}{lrl} and \#\textcolor{gray}{lrr} that contain, respectively 2 and 1 differences. In both cases $\rec$ succeeds. 
At this stage, all the child partitions of \#\textcolor{gray}{l} have been recovered.
Let us now go back to partition \#\textcolor{gray}{r}, which gets split into partitions \#\textcolor{gray}{rl} and \#\textcolor{gray}{rr} that contain, respectively, 0 and 4 differences. 
The call to $\rec$ associated with partition \#\textcolor{gray}{rl} succeeds, but unfortunately, it does not recover any difference since the partition was empty.
The call to $\rec$ in partition \#\textcolor{gray}{rr} fails, and the partition is further split into partitions \#\textcolor{gray}{rrl} and \#\textcolor{gray}{rrr}, both of which contain 2 differences. Thus, the call to $\rec$ succeeds in both cases. At this stage, all the child partitions of \#\textcolor{gray}{r} have been processed. Hence, the sets are reconciled and  \ac{PSR} finalizes.
\begin{figure}[t]
	\centering
	\scalebox{0.7}{
		\scalebox{0.96}{
\begin{tikzpicture}
   [level 0/.style = {sibling distance=10.5cm},
   level 1/.style = {sibling distance=5.0cm},
   level 2/.style = {sibling distance=2.8cm},
   level 3/.style = {sibling distance=2.2cm}, level distance = 1.5cm , minimum size=1cm]

   \node [circle, draw, align=center] (root){\redX \\ $9$}
   child {node [circle, draw, align=center] (l) { \redX \\ $5$}
      child {node[circle, draw, align=center] (ll) { \greencheck \\ $2$}         
      }
      child {node [circle, draw, align=center] (lr) { \redX \\ $3$}
           child {node[circle, draw, align=center] (lrl) { \greencheck \\ $2$}}
         child {node[circle, draw, align=center] (lrr) { \greencheck \\ $1$}}
      }
   }
   child {node [circle, draw, align=center] (r) { \redX \\ $4$}
       child {node[circle, draw, align=center] (rl) { \greencheck \\ $0$}         
      }
       child {node[circle, draw, align=center] (rr) { \redX \\ $4$}               
           child {node[circle, draw, align=center] (rrl) { \greencheck \\ $2$}}
         child {node[circle, draw, align=center] (rrr) { \greencheck \\ $2$}}
       }
      };

\node[above left= -0.4cm and -0.1cm of root] ( ) {\#\textcolor{gray}{\#root}};
\node[above left= -0.7cm and -0.1cm of l] ( ) {\#\textcolor{gray}{l}};
\node[above left= -0.7cm and -0.1cm of ll] ( ) {\#\textcolor{gray}{ll}};
\node[above left= -0.7cm and -0.1cm of lr] ( ) {\#\textcolor{gray}{lr}};
\node[above left= -0.6cm and -0.1cm of lrl] ( ) {\#\textcolor{gray}{lrl}};
\node[above left= -0.6cm and -0.1cm of lrr] ( ) {\#\textcolor{gray}{lrr}};
\node[above left= -0.7cm and -0.1cm of r] ( ) {\#\textcolor{gray}{r}};
\node[above left= -0.6cm and -0.1cm of rl] ( ) {\#\textcolor{gray}{rl}};
\node[above left= -0.6cm and -0.1cm of rr] ( ) {\#\textcolor{gray}{rr}};
\node[above left= -0.6cm and -0.1cm of rrl] ( ) {\#\textcolor{gray}{rrl}};
\node[above left= -0.6cm and -0.1cm of rrr] ( ) {\#\textcolor{gray}{rrr}};

\node at ($ (root) + (6.7cm,0) $) (a0) {level 0};
\node at  (l -| a0)(b0) (b0) {level 1};
\node at  (ll -| a0)(b0) (b0) {level 2};
\node at  (lrl -| a0)(b0) (b0) {level 3};

\end{tikzpicture}
}
	}
	\caption{Partition tree associated with \acs{PSR} with $\bardelta=2$. Each node represents a partition. The number inside each node represents the number of differences in the associated partition.
    Nodes marked with {\footnotesize \redX} are those in which $\rec$ fails because they contain more than $\bardelta=2$ differences. Nodes marked with {\footnotesize \greencheck} are those for which the $\rec$ succeeds because they contain at most $\bardelta=2$ differences.     
    The number outside each node marked in gray corresponds to its associated label (\#).     
    }
	\label{fig:tree_no_sic_m_2}
\end{figure}

\end{example}
In this example, \acs{PSR} needed to call $\rec$ a total of $11$ times. Hence, the communication cost is $11$ times the size of the \ac{SR} data structure, and its time complexity will be dominated by the corresponding $11$ calls to $\rec$.
We now examine the round complexity of \ac{PSR} in this example. The number of communication rounds equals the number of levels along the longest root-to-leaf path in the partition tree. Consider one  deepest leaf, e.g., the node labeled \#\textcolor{gray}{rrr}. The path from the root to this node visits four nodes (\#\textcolor{gray}{root}, \#\textcolor{gray}{r}, \#\textcolor{gray}{rr}, \#\textcolor{gray}{rrr}). Each of these 4 nodes corresponds to one communication round (one request from node~A followed by a reply from node~B). Hence, in this example a total of 4 communication rounds occur.

    \section{Enhanced Partitioned Set Reconciliation } \label{sec:enhanced}

\ac{EPSR} draws inspiration from concepts found in tree algorithms utilized in random-access protocols \cite{capetanakis1979tree,yu2007high,stefanovic2023tree,Flajolet,SGDK2020}. It specifically capitalizes on the mechanism known as successive interference cancellation, commonly employed in communication channels. Within the framework of set reconciliation, this approach involves employing the subtract operation of the \ac{SR} data structure to remove the differences contained in the child partition from the parent one.

Algorithms~\ref{alg:EPSR} and \ref{alg:EPSR_r} provide the \ac{EPSR} pseudocode. The initial call (Alg.~\ref{alg:EPSR}) invokes the recursive routine in Algorithm~\ref{alg:EPSR_r} with \texttt{skip}$=$\texttt{false}.
Similarly to \ac{PSR}, for every partition, \ac{EPSR} first calls $\rec$. If $\rec$ succeeds, the algorithm terminates. However, if $\rec$ fails, the sets are partitioned, and the resulting partitions are processed. The key difference lies in how these partitions are processed compared to \ac{PSR}. Specifically:

\begin{itemize}
\item In \ac{EPSR} we retain the \ac{SR} data structure of the current partition, denoted as $\Z$.

\item After receiving $B$'s \ac{SR} data structure (line 10) for the $i$-th partition, and computing the data structure $\Z_i$ that contains the differences of the $i$-th partition, this data structure is subtracted from the parent data structure $\Z$. This yields a data structure that contains all differences in the partitions $i+1$ to $c$. Next, recovery is attempted in parallel for $\Z_i$ and $\Z$.

\item 
The processing of the $c$-th partition differs from previous partitions. Specifically, when \ac{EPSR} invokes itself to process the $c$-th partition (line 20), the flag $\flagskip$ is set to \emph{true}. This ensures that we skip transmitting the \ac{SR} data structure of the $c$-th partition since it is already available. Additionally, we also skip the call to $\rec$, as it is guaranteed to fail.\footnote{After processing the $c-1$-th partition and deleting the recovered differences from $\Z$, the recovery operation of $\Z$ has already been attempted and failed; otherwise, the algorithm would not proceed to process the $c$-th partition. At this point, all the remaining differences in $\Z$ are contained within the $c$-th partition, meaning that $\Z$ and $\Z_c$ are identical, and their number of differences is guaranteed to be strictly larger than $\bardelta$, since otherwise the algorithm would have ended on line 19. Consequently, if $\rec$ has failed for $\Z$, it will also fail for $\Z_c$.} 

\end{itemize}

\begin{algorithm}[t]
\caption{$[\D] = \textsc{EPSR}(\SA,\P)$}
\label{alg:EPSR}

\Input{
  $\SA$ — Host A’s set\;
  $\SB$ — Host B’s set\;
  $\P$  — partition to process
}
\Output{$\D = (\SA\cap\P)\ominus(\SB\cap\P)$}

\BlankLine

$\Z_A.\mathsf{insert\_set}(\SA \cap \P)$\;
$\Z_B \gets \req(\SB \cap \P)$ \tcp*[r]{remote @ B}
$\Z_0 \gets \mathsf{subtract}(\Z_A,\Z_B)$\;

$[\D] \gets \textsc{EPSR\_r}(\SA,\P,\Z_0,\texttt{false})$\;

\Return{$\D$}\;
\end{algorithm}

\begin{algorithm}[t] 
\caption{$[\D] = \textsc{EPSR\_r}(\SA,\P,\Zp,\flagskip)$} %
\label{alg:EPSR_r}
\Input{

  $\SA$ — Host A’s set\;  
  $\P$  — partition to process\;
  $\Zp$ — initial data SR structure\;
  $\flagskip$ — skip flag
}
\Output{
  $\D = (\SA\cap\P)\ominus(\SB\cap\P)$
}

\BlankLine
$\D \gets \emptyset$\;
$\Z \gets \Zp$\;

\If{not $\flagskip$}{
    $[\flag,\D] \gets \Z.\mathsf{recovery}()$\;
    \If{$\flag$}{\Return{$[\D]$}}
}

$\{\P_1,\dots,\P_c\} \gets \mathsf{Partition}(\P)$\;

\For{$i \gets 1$ \KwTo $c-1$}{%
    $\Z_{A_i}.\mathsf{insert\_set}(\SA \cap \P_i)$\;
    $\Z_{B_i} \gets \req(\SB \cap \P_i)$\;
    $\Z_i \gets \mathsf{subtract}(\Z_{A_i},\Z_{B_i})$\;

    \Spawn{        
        $[\D_i] \gets
           \textsc{EPSR\_r}(\SA,\P_i,\Z_i,\texttt{false})$\;       
        $\D \gets \D \cup \D_i$ \tcp*[r]{atomic }
    } 

    $\Z \gets \mathsf{subtract}(\Z,\Z_i)$\;
    $[\flag',\D_c] \gets \Z.\mathsf{recovery}()$\;
    \If{$\flag'$}{        
        $\D \gets \D \cup \D_c$ \tcp*[r]{atomic }
        \WaitAllChildren\tcp*[r]{wait for the spawned child to finish}
        \Return{$[\D]$}
    }
} 

$[\D_c] \gets \textsc{EPSR\_r}(\SA,\P_c,\Z,\texttt{true})$\;
$\D \gets \D \cup \D_c$\tcp*[r]{atomic}

\WaitAllChildren\tcp*[r]{wait for all spawned children to finish}

\Return{$[\D]$}\;
\end{algorithm}

\begin{figure}[t]
	\centering
	\scalebox{0.7}{
		\scalebox{0.96}{

\begin{tikzpicture}
   [level 0/.style = {sibling distance=10.5cm},
   level 1/.style = {sibling distance=5.0cm},
   level 2/.style = {sibling distance=2.8cm},
   level 3/.style = {sibling distance=2.2cm}, level distance = 1.5cm , minimum size=1cm]

   \node [circle, draw, align=center] (root){\redX \\ $9$}
   child {node [circle, draw, align=center] (l) { \redX \\ $5$}
      child {node[circle, draw, align=center] (ll) { \greencheck \\ $2$}         
      }
      child {node [circle, draw,  fill=white,
      postaction={pattern={Lines[angle=90,distance={6pt},line width=3pt]}, pattern color=black!20},  align=center] (lr) { \redX \\ $3$}
           child {node[circle, draw, align=center] (lrl) { \greencheck \\ $2$}}
         child {node[circle, draw, fill=white,
      postaction={pattern={Lines[angle=0,distance={6pt},line width=3pt]}, pattern color=black!20}, dashed, align=center] (lrr) { \greencheck \\ $1$}}
      }
   }
   child {node [circle, draw, fill=white,
      postaction={pattern={Lines[angle=90,distance={6pt},line width=3pt]}, pattern color=black!20},  align=center] (r) { \redX \\ $4$}
       child {node[circle, draw, align=center] (rl) { \greencheck \\ $0$}         
      }
       child {node[circle, draw, fill=white,
      postaction={pattern={Lines[angle=90,distance={6pt},line width=3pt]}, pattern color=black!20},  align=center] (rr) { \redX \\ $4$}               
           child {node[circle, draw, align=center] (rrl) { \greencheck \\ $2$}}
         child {node[circle, draw,  fill=white,
      postaction={pattern={Lines[angle=0,distance={6pt},line width=3pt]}, pattern color=black!20}, dashed,  align=center] (rrr) { \greencheck \\ $2$}}
       }
      };

\node[above left= -0.5cm and -0.1cm of root] ( ) {\#\textcolor{gray}{root}};
\node[above left= -0.7cm and -0.1cm of l] ( ) {\#\textcolor{gray}{l}};
\node[above left= -0.7cm and -0.1cm of ll] ( ) {\#\textcolor{gray}{ll}};
\node[above left= -0.7cm and -0.1cm of lr] ( ) {\#\textcolor{gray}{lr}};
\node[above left= -0.6cm and -0.1cm of lrl] ( ) {\#\textcolor{gray}{lrl}};
\node[above left= -0.6cm and -0.1cm of lrr] ( ) {\#\textcolor{gray}{lrr}};
\node[above left= -0.7cm and -0.1cm of r] ( ) {\#\textcolor{gray}{r}};
\node[above left= -0.7cm and -0.1cm of rl] ( ) {\#\textcolor{gray}{rl}};
\node[above left= -0.7cm and -0.1cm of rr] ( ) {\#\textcolor{gray}{rr}};
\node[above left= -0.6cm and -0.1cm of rrl] ( ) {\#\textcolor{gray}{rrl}};
\node[above left= -0.6cm and -0.1cm of rrr] ( ) {\#\textcolor{gray}{rrr}};

\node at ($ (root) + (6.7cm,0) $) (a0) {level 0};
\node at  (l -| a0)(b0) (b0) {level 1};
\node at  (ll -| a0)(b0) (b0) {level 2};
\node at  (lrl -| a0)(b0) (b0) {level 3};

\end{tikzpicture}
}
	}
	\caption{Partition tree associated with \ac{EPSR} with $\bardelta=2$. Each node represents a partition. The number inside each node represents the number of differences in the associated partition.
    Nodes marked with {\footnotesize \redX} are those in which the \rec fails because they contain more than $\bardelta=2$ differences. Nodes marked with {\footnotesize \greencheck} are those for which \rec succeeds because they contain at most $\bardelta=2$ differences.  
    Nodes drawn with a solid line and filled with a vertical line pattern are those associated with partitions where \ac{EPSR} was invoked with $\flagskip=\text{true}$ (line 20 in Algorithm~\ref{alg:EPSR_r}). The \ac{EPSR} calls associated with these nodes skip the request to B for a data structure (and B's response). Hence, these nodes do not contribute to the time and communication cost.
    Similarly, the nodes drawn with a dashed line and filled with a horizontal line pattern are those associated with partitions for which \ac{EPSR} is not invoked since reconciliation succeeds in the parent partition.    
    The number outside each node marked in gray corresponds to its associated label (\#).}
	\label{fig:tree_sic_m_2}
\end{figure}

\begin{example}[\Acf{EPSR}]\label{example:EPSR}
Consider the example shown in Fig.~\ref{fig:tree_sic_m_2}, where \acs{EPSR} is instantiated with $\bar m=2$ and $c=2$, and where, for comparison, the partitioning is identical to that in Example~\ref{example:PSR}.
In contrast to \ac{PSR}, in \ac{EPSR} after each split only one \ac{SR} data structure is transmitted: that of the left child. The \ac{SR} for the right child is obtained locally by subtracting the left child's \ac{SR} from the parent's \ac{SR}. Nodes whose \ac{SR} is transmitted over the network are drawn with a white fill, whereas nodes whose \ac{SR} is obtained via the \texttt{subtract} operation are shown with vertical lines.
At level~0 the partition \#\textcolor{gray}{root} contains $9$ differences, so recovery fails and \#\textcolor{gray}{root} is split into \#\textcolor{gray}{l} and \#\textcolor{gray}{r} (level~1). The \ac{SR} of \#\textcolor{gray}{l} is transmitted. The \ac{SR} of \#\textcolor{gray}{r} is computed by subtraction. Both partitions are then processed concurrently. Since both contain more than $\bar m=2$ differences, recovery fails for each and they are split again.
Focus first on \#\textcolor{gray}{l}, which splits into \#\textcolor{gray}{ll} and \#\textcolor{gray}{lr}. Again, only the \ac{SR} of \#\textcolor{gray}{ll} is transmitted; the \ac{SR} of \#\textcolor{gray}{lr} is computed. Both children are processed in parallel. Partition \#\textcolor{gray}{ll} contains $2$ differences, so recovery succeeds and two elements are recovered. Partition \#\textcolor{gray}{lr} contains $3$ differences, so recovery fails and it is split into \#\textcolor{gray}{lrl} and \#\textcolor{gray}{lrr} (level~3). The \ac{SR} of \#\textcolor{gray}{lrl} is transmitted; the \ac{SR} of \#\textcolor{gray}{lrr} is computed. Both now contain at most two differences, so recovery succeeds in each, recovering $2$ and $1$ elements, respectively.
Let us return to \#\textcolor{gray}{r}, which contains $4$ differences and is split into \#\textcolor{gray}{rl} and \#\textcolor{gray}{rr}. The \ac{SR} of \#\textcolor{gray}{rl} is transmitted.  The \ac{SR} of \#\textcolor{gray}{rr} is computed. Both are processed concurrently. Partition \#\textcolor{gray}{rl} contains $0$ differences, so recovery succeeds (although no differences are recovered). Partition \#\textcolor{gray}{rr} contains $4$ differences and is split into \#\textcolor{gray}{rrl} and \#\textcolor{gray}{rrr}. The \ac{SR} of \#\textcolor{gray}{rrl} is transmitted. The \ac{SR} of \#\textcolor{gray}{rrr} is computed. Both contain $2$ differences, so recovery succeeds for each, and the sets are reconciled.
\end{example}

In this example, \ac{EPSR} needed to transmit $6$ data structures, one for each node shown filled in solid white, compared to the $11$ needed by \ac{PSR}.
Regarding the time complexity, we have that every node filled in solid white is associated with an invocation of \emph{recovery}, and thus with a call to the \emph{subtract} and \emph{recovery} operations. 
Thus, \ac{EPSR} calls \emph{subtract} and \emph{recovery} $11$ times, the same as \ac{PSR}.
We now examine the round complexity of \ac{EPSR} in this example. For $c=2$, at each split \ac{EPSR} reuses the parent \ac{SR} for one child and requests a single new \ac{SR} for the other. The two children can therefore be processed within the same round. Consequently, the number of communication rounds equals the length of the longest root-to-leaf path in the partition tree, as in \ac{PSR}. Consider a deepest leaf, e.g., the node labeled \#\textcolor{gray}{rrr}. The path from the root to this node visits 4 nodes (\#\textcolor{gray}{root}, \#\textcolor{gray}{r}, \#\textcolor{gray}{rr}, \#\textcolor{gray}{rrr}). Each node corresponds to one communication round (one request from node~A followed by a reply from node~B). Hence, in this example a total of 4 communication rounds occur.

This example shows that, for $c=2$,  the communication cost of \ac{EPSR} is lower than that of \ac{PSR}, while keeping the same round complexity.

    \section{Analysis}\label{sec:analysis}

In this section, we provide recursive formulas for the  expected   communication cost,  time complexity, and number of communication rounds for both \ac{PSR} and \ac{EPSR},  given the (initial) number of differences $\d$ between two sets $\SA$ and $\SB$.
The communication cost of \ac{PSR} and \ac{EPSR} in Algorithms~\ref{alg:psr} and \ref{alg:EPSR_r} is associated with the number of requests sent to node B to obtain the \ac{SR} data structure of a partition. 
Regarding time complexity, as shown in Table~\ref{tab:complexity}, the \rec operation 
has a complexity of $\mathcal{O}(\bardelta^3)$, whereas all other operations exhibit linear 
complexity in $\bardelta$. Consequently, already for moderate values of  $\bardelta$, the time complexity 
of the algorithms will primarily be determined by the number of calls to the \rec function. 
Thus, we measure the time complexity in terms of the number of \rec function calls.

In our analysis, we assume that the elements of $\SA$ and $\SB$ are chosen uniformly at random from a universal set $\mathcal{U}$. 
Consider  a set element $s$ chosen uniformly at random in $\mathcal{U}$ and let $\P \subseteq \mathcal{U}$ be partitioned into $c$ disjoint subsets
$\{\P_1, \P_2, \dots, \P_c\}$ with $\P = \biguplus_{j=1}^c \P_j$.
Conditional on $s \in \P$, the probability that $s$ falls into the $j$-th subset is
\[
p_j \coloneq  \frac{|\P_j|}{\sum_{i=1}^c |\P_i|}=\frac{|\P_j|}{|\P|}.
\]

\subsection{PSR algorithm}

Examining Algorithm~\ref{alg:psr}, we observe that in \ac{PSR}, the count of \rec calls and the number of \ac{SR} data structures exchanged over the network match, since every request to B for a \ac{SR} data structure is followed by a call to \rec.

Let us denote by $\npc_\d $ the random variable associated with the number of calls to \rec when executing \ac{PSR} (Algorithm~\ref{alg:psr}), given $\d$.
Note that $\npc_\d$ depends implicitly on $\bardelta$ and on the partitioning schedule, that is, on $p_j$.

\begin{theorem}
\label{thm:PSR}
The expected number of \rec calls in the \ac{PSR} algorithm, $\expnpc_\d \coloneq \mathrm{E}[\npc_\delta]$, is governed by the following recursive expression
\begin{align}
\label{eq:expnpc}
    \expnpc_\d = \begin{cases}
        1 & 0 \leq \d \leq \bardelta \\
        \frac{1 + \sum_{j=1}^{c} \sum_{i = 0}^{\d - 1} {\d \choose i } p_j^{i} (1 - p_j)^{\d - i} \expnpc_{i}}{1 - \sum_{j = 1}^{c} p_j^\d} & \text{otherwise.}
    \end{cases} 
\end{align}
\end{theorem}
The proof of Theorem~\ref{thm:PSR} is provided in Appendix~\ref{app:PSR}.
One relevant special case of Theorem~\ref{thm:PSR} is that of \emph{fair} partitioning, that is, $p_j =1/c $, $\forall j$, for which the expected number of \rec calls is given in the following corollary.

\begin{corollary}
\label{cor:fair}
When fair partitioning is used, $p_j =1/c $, $\forall j$,  the expected number of \rec calls in the \ac{PSR} algorithm  follows the recursive expression
\begin{align}
\label{eq:expnpcfair}
    \expnpc_\d = \begin{cases}
       1 & 0 \leq \d \leq \bardelta \\
        \frac{1 + \frac{1}{c^{\d - 1}} \sum_{i = 0}^{\d - 1} {\d \choose i} (c - 1)^{\d - i} \expnpc_{i}}{1 - \frac{1}{c^{\d - 1}}} & \text{otherwise.}
    \end{cases} 
\end{align}
\end{corollary}

Theorem~\ref{thm:PSR} provides an exact finite length analysis of the communication cost of \ac{PSR} for arbitrary partitioning. The following lemma, borrowed from \cite[Theorem 3]{Minsky2002},  provides a simple yet tight bound for the expected communication cost of \ac{PSR} when fair\footnote{Deriving precise bounds for arbitrary partitioning exceeds the scope of the current work. We plan to address this in a separate study.} partitioning is used,
\begin{lemma}
\label{lem:psr}
When fair partitioning is used, $p_j =1/c $, $\forall j$, the expected number of calls to \rec in the \ac{PSR} algorithm is at most \cite[Theorem 3]{Minsky2002}
\[
\expnpc_\d \leq 8 e (c+1) \frac{\delta}{\bardelta +1 }
\]
\end{lemma}

Let us now look into the number of communication rounds required by \ac{PSR}. A prominent feature of \ac{PSR} is that, after partitioning a set, its $c$ child partitions can be processed in parallel since there is no interdependency between them. Thus, the number of communication rounds corresponds to the depth of the partition tree, i.e., the number of levels. The following lemma provides the expected communication-round complexity of \ac{PSR} for arbitrary partitioning.
\begin{lemma}
\label{lem:psr_round}
The expected number of communication rounds in the \ac{PSR} algorithm is
\[
 \mathcal{O}\left( \lambda \log \left( \frac{\d}{\bardelta}\right) \right) 
\]
where $\lambda = - \frac{1}{\log(\pmax)}$,  and $\pmax=\max_{i=1,\dots,c} p_i$.
\end{lemma}
The proof of Lemma~\ref{lem:psr_round} can be found in Appendix~\ref{app:psr_round}.
For fair partitioning, the result in Lemma~\ref{lem:psr_round} particularizes to  $\mathcal{O}(\log_c \left( \d/ \bardelta\right) )$. 

\subsection{EPSR algorithm}
\label{sec:EPSR}
From the 
definition of 
Algorithms~\ref{alg:EPSR} and \ref{alg:EPSR_r}, it is evident that the number of \ac{SR} data structures sent over the network and the number of \emph{recovery} calls differ in \ac{EPSR}. The \emph{recovery} function is invoked once after receiving each \ac{SR} data structure and additionally at line 16 of Algorithm \ref{alg:EPSR_r}. Therefore, it is necessary to analyze  the number of \ac{SR} data structures exchanged and the number of \emph{recovery} calls separately.

Let $\npce_\d$ represent the random variable associated with the number of \ac{SR} data structures exchanged in \ac{EPSR} (Algorithm~\ref{alg:EPSR_r}), given $\d$. Note that $\npce_\d$ implicitly depends on $\bardelta$ and the partitioning schedule (specifically on $p_j$).

\begin{theorem}
\label{thm:EPSR}
The expected number of \ac{SR} data structures exchanged in \ac{EPSR} algorithm, $\expnpce_\d \coloneq \mathrm{E}[\npce_\d]$, is $\expnpce_\d=1$ when $0 \leq \d \leq \bardelta$ and otherwise it corresponds to
\begin{align}
    \expnpce_\d  = &\frac{\sum_{j=1}^{c} \sum_{i = 0}^{\d - 1} {\d \choose i } p_j ^{i} ( 1 - p_j )^{\d - i } \expnpce_{i}  }{ 1 - \sum_{j=1}^c p_j^\d}   \\
    - &  \frac{ \sum_{j=1}^{c-2} \sum_{i=0}^\bardelta \binom{\d}{i} (1 - F_j)^i F_j^{\d - i}}{1 - \sum_{j=1}^c p_j^\d}
\end{align}
where $F_j = \sum_{i = 1}^{j} p_i$.
\end{theorem}
The proof of Theorem~\ref{thm:EPSR} is deferred to  Appendix~\ref{app:EPSR}.

Let us now look into the time complexity. Let $\nrec_\d$ denote the random variable associated with the number of calls to the \emph{recovery} operation, which implicitly depends on $\bardelta$ and $p_j$.

\begin{theorem}
\label{thm:EPSRrec}
The expected number of \emph{recovery} calls in the \ac{EPSR} algorithm, $\expnrec_\d \coloneq \mathrm{E}[\nrec_\d]$, is $\expnrec_\d=1$ when $0 \leq \d \leq \bardelta$ and otherwise it corresponds to
\begin{align}
    \expnrec_\d  & =\frac{\sum_{j=1}^{c} \sum_{i = 0}^{\d - 1} {\d \choose i } p_j ^{i} ( 1 - p_j )^{\d - i } \expnrec_{i}  }{ 1 - \sum_{j=1}^c p_j^\d}   \\
    & \phantom{=} +  \frac{ (c - 1) - 2 \sum_{j=1}^{c-2} \sum_{i=0}^\bardelta \binom{\d}{i} (1 - F_j)^i F_j^{\d - i}}{1 - \sum_{j=1}^c p_j^\d}. 
\end{align}
where $F_j = \sum_{i = 1}^{j} p_i$.
\end{theorem}
The proof of Theorem~\ref{thm:EPSRrec} is provided in Appendix~\ref{app:EPSR_rec}.

The following lemma characterizes the number of communication rounds required by \ac{EPSR}. 
\begin{lemma} 
\label{lemma:epsr_rounds}
The expected number of communication rounds in the \ac{EPSR} algorithm is
\[
 \mathcal{O}\left( \lambda^\star \log \left( \frac{\d}{\bardelta}\right) \right) 
\]
where 
\[\lambda^\star = - \frac{1}{\log(\qmax)}, \qquad
{\qmax = \max_i q_i}\] 
\[{q_i = \max \left( p_i^\star, 1-p_i^\star\right)}, \quad 
{p_i^\star= \frac{p_i}{1-\sum_{j=1}^{i-1} p_j}}.\]
\end{lemma}
The proof of Lemma~\ref{lemma:epsr_rounds} can be found in  Appendix~\ref{app:EPSR_rounds}.

It is easy to see that the choice of $p_i$ that minimizes the number of rounds is that which yields $q_i = 1/2$, which implies $p_i^\star = 1/2$. Therefore, we have
\begin{align} \label{eq:p_i_val}
    p_j = \begin{cases}
        2^{-j}       & j=1,\dots, c-1 \\
        2^{-(c-1)}   & j=c. 
    \end{cases}
\end{align}
For this particular choice of $p_i$, the expected number of communication rounds is 
 $\mathcal{O}(\log_2 \left( \d/ \bardelta\right) )$.
 Two observations can be made here. The first one is that for this choice of the partitioning probabilities $p_i$, the number of communication rounds for \ac{EPSR} shows no dependency on $c$. 
 The second one is that the number of communication rounds for \ac{EPSR} is equivalent to that of binary \ac{PSR}.

    \section{Numerical Results}\label{sec:num_res}

\subsection{Verification}\label{sec:verif}

\begin{figure}[t]
    \centering    
        \includegraphics[width=0.99\columnwidth]{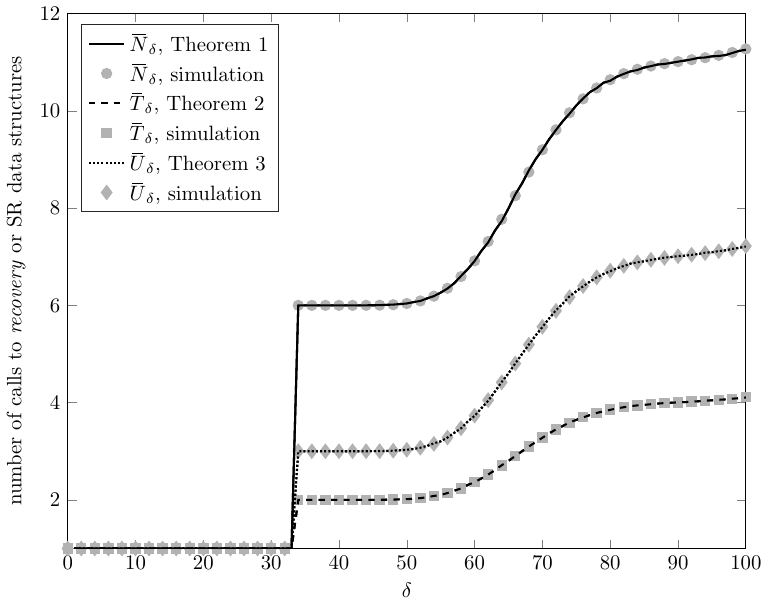}      
    \caption{Number of calls to  \rec ($\expnpc_\d$, $\expnrec_\d$) and \ac{SR} data structures transmitted ($\expnpce_\d$) as a function of the number of differences $\d$. The parameters considered are $\bardelta=33$, $c=5$, $p_1= 0.15$, $p_2=0.1$, $p_3=0.25$, $p_4=0.2$, and $p_5=0.3$.}
    \label{fig:example}
\end{figure}

Fig.~\ref{fig:example} compares the analytical results derived in Theorems~\ref{thm:PSR}, \ref{thm:EPSR}, and \ref{thm:EPSRrec} with corresponding results obtained from Monte Carlo simulations.  An arbitrary parameter set was chosen for this illustrative comparison: $\bardelta=33$, $c=5$, $p_1=0.15$, $p_2=0.1$, $p_3=0.25$, $p_4=0.2$, and $p_5=0.3$. For each value of $\d$ considered, $10^4$ independent samples (random partitions of the $\d$ differences) were generated for both the \ac{PSR} and \ac{EPSR} algorithms.  
The simulation results show excellent alignment with the analytical predictions. We conducted simulations with numerous other parameter values as well, and consistently achieved an excellent match with our analytical predictions.

\subsection{Evaluation}\label{sec:eval}

In this section, we provide numerical results for the following performance metrics, which are derived from the expected communication cost and time complexity obtained in the previous section. 

\begin{definition}
\emph{Redundancy} is defined as the communication cost (measured in number of bits) normalized by the size of the set difference, $\d \ell$.
\end{definition}
The minimum (optimal) value for the redundancy of a set reconciliation protocol is one. In fact, \ac{CPI} achieves a redundancy only slightly greater than 1, albeit at the price of a very high time complexity.
Thus, from Eq.\eqref{eq:cost} we have that for \ac{PSR} the redundancy is $\expnpc_\d \left( (\bardelta+\kfact+1) (\ell +1) -1 \right) / \left( \d \ell \right)$, whereas for \ac{EPSR} it is $\expnpce_\d \left( (\bardelta+\kfact+1) (\ell +1) -1 \right) / \left( \d \ell \right)$.

\begin{definition}
\emph{Normalized complexity} is defined as the expected number of \rec operation calls divided by $\d / \bardelta$. 
\end{definition}
Consequently, if a protocol exhibits a normalized complexity of, for example, 4, it indicates that, on average, 4 calls to \rec are required to recover $\bardelta$ differences. Therefore, the optimal (minimum) achievable normalized complexity for a set reconciliation protocol is one, meaning each call to \rec successfully retrieves $\bardelta$ differences, which is the maximum possible.
Hence, we have that the normalized complexity of \ac{PSR} is $\frac{\expnpc_\d \bardelta}{\d} $ whereas for \ac{EPSR} it is given by $\frac{ \expnrec_\d \, \bardelta} {\d} $.

\subsubsection{Parameter Selection for EPSR} \label{sec:param} 

~

Let us focus first on choosing an appropriate partition strategy for 
\ac{EPSR}. Specifically, the choice involves selecting the partition factor $c$ and the partition probabilities $p_j$, for $j=1, \dots, c$.
In \cite{VDSK2024}, in the context of tree algorithms for random access protocols, the optimal partition probabilities were studied for the case of $\bardelta= 1$ in the asymptotic regime when $\delta$ tends to infinity. In particular, it was found out that 
choosing $p_j$ according to \eqref{eq:p_i_val}, not only minimizes the number of communication rounds but also  minimizes $\lim_{\d \rightarrow \infty} \expnpc_\d$.
Although the proof in \cite{VDSK2024} was derived just for the case that $\bardelta= 1$, our numerical results suggest that these partition probabilities remain optimal\footnote{Specifically, we carried out numerical optimization for finite yet large values of $\d$ and always obtained partition probabilities nearly identical to Eq.~\eqref{eq:p_i_val}.} for general values of $\bardelta$.\footnote{Further considerations of this matter are beyond the scope of this work and are left for further study.} 
In the rest of this section we will always assume that $p_j$ is given by Eq. \eqref{eq:p_i_val}.

\begin{figure}[t]
    \centering
    \subfloat[redundancy\label{fig:EPSR_c_cost}]{
        \includegraphics[width=0.99\columnwidth]{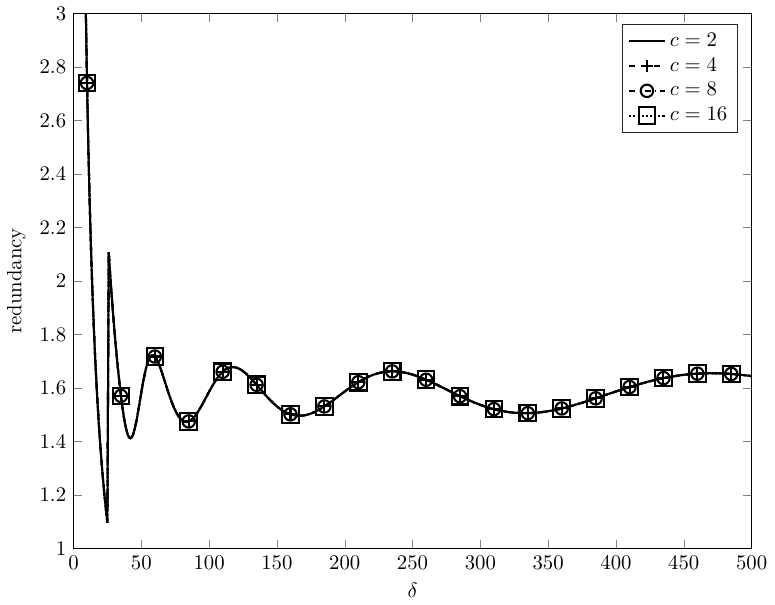}}  
        
    \subfloat[normalized complexity\label{fig:EPSR_c_complexity}]{
        \includegraphics[width=0.99\columnwidth]{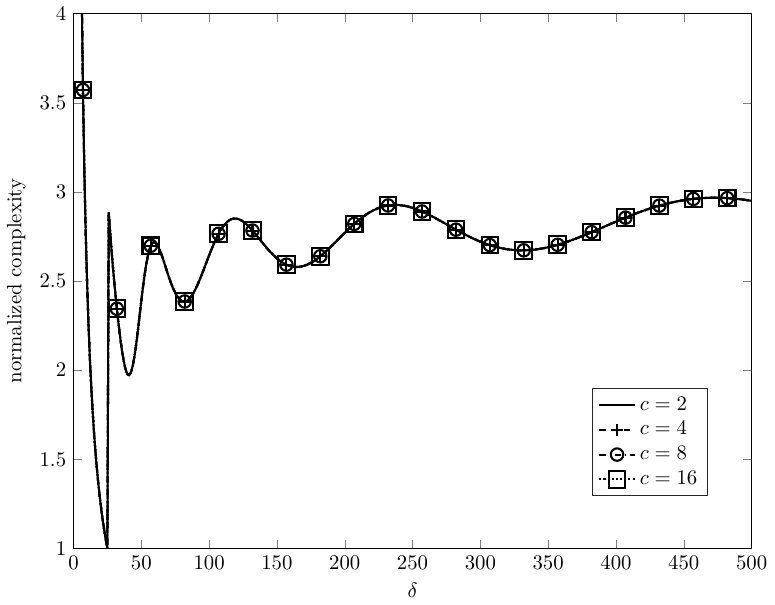}}
    \caption{Performance as a function of the number of differences $\d$ for \ac{EPSR} with $c=(2, 4, 8, 16)$. In all cases we assume $\kfact =1$,  $\ell=64$ and $\bardelta=25$ as parameters of the \ac{SR} data structure. The upper figure (a) shows the redundancy whereas the lower figure (b) shows the normalized complexity.}
    \label{fig:EPSR_c}
\end{figure}

The lingering question is how to select the partition factor~$c$. In Fig.~\ref{fig:EPSR_c} we show the redundancy and normalized complexity as a function of $\d$ for $c=2, 4, 8,$ and $16$. Looking at the figure, it is apparent that, when $p_j$ follows Eq.~\eqref{eq:p_i_val}, the redundancy (i.e., communication cost) and normalized complexity (i.e., time complexity) remain the same\footnote{In the figure, we have focused solely on values of $c$ that are powers of $2$. Nonetheless, the pattern holds for general values of $c$: redundancy and normalized complexity do not depend on $c$.} independently of the value of $c$. 

For the sake of simplicity, in the rest of this Section we will set $c=2$. Interestingly, for $c=2$, Eq.~\eqref{eq:p_i_val} yields $p_1=p_2=1/2$, i.e., fair  partitioning.

Now that we have narrowed down what is a good partition strategy, the remaining question is which value to choose for $\bardelta$. 
\begin{figure}[t]
    \centering
    \includegraphics[width=0.99\columnwidth]{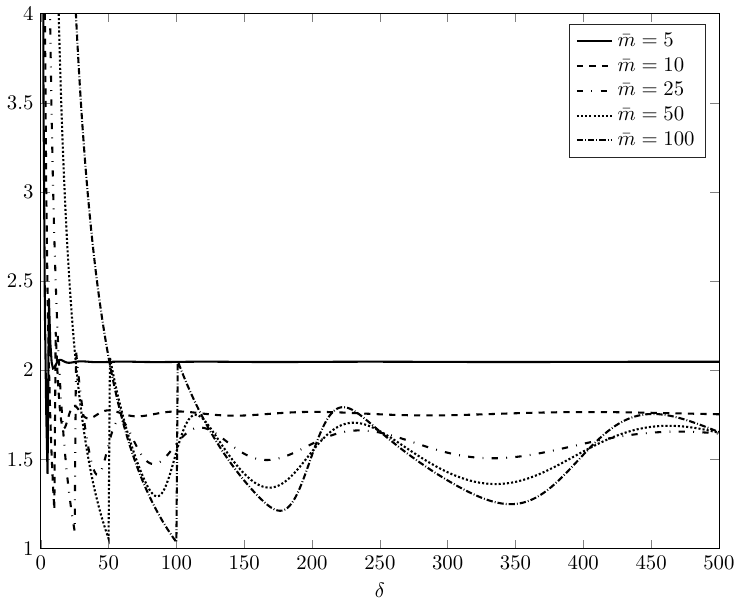}
    \caption{EPSR redundancy versus number of differences $\d$ under $\bardelta$ values of 5, 10, 25, 50 and 100, assuming $c=2$, $\kfact=1$, $\ell=64$, and fair partitioning, $p_1=p_2=1/2$.}
    \label{fig:cost_vs_delta_unequal}
\end{figure}
In Fig.~\ref{fig:cost_vs_delta_unequal}, we present the redundancy of \ac{EPSR} as a function of the number of differences $\d$ for various values of $\bardelta$: 5, 10, 25, and 100. We observe that, except for small $\d$ values, the redundancy fluctuates around a specific level. As $\bardelta$ increases, redundancy generally decreases, but the fluctuations become more pronounced.

Given this trend, a larger $\bardelta$ may seem advantageous. However, this comes at the cost of increased computational complexity, as the complexity of the \rec operation scales as $\mathcal{O}(\bardelta^3)$. Therefore, choosing a moderate value (e.g., $\bardelta=25$ or $\bardelta=50$ ) offers a practical trade-off between minimizing redundancy and keeping computation manageable.

The oscillatory behavior in Figs.~\ref{fig:EPSR_c} and \ref{fig:cost_vs_delta_unequal} can also be observed in the remaining figures in the paper.
This behavior is characteristic for tree algorithms, and was also noticed in, e.g., \cite{Flajolet, stefanovic2023tree}, and is a consequence of the discrete number of tree levels that arise during the partition process \cite{jong}.
A characterization of the periodicity of the oscillations in terms of $\bardelta$, $c$, partition probabilities, and $\d$ is beyond the scope of this paper and is left for further investigation.

\subsubsection{Comparison with PSR}  ~

\begin{figure}[t]
    \centering
    \subfloat[redundancy \label{fig:cost_vs_delta_comp}]{
        \includegraphics[width=0.99\columnwidth]{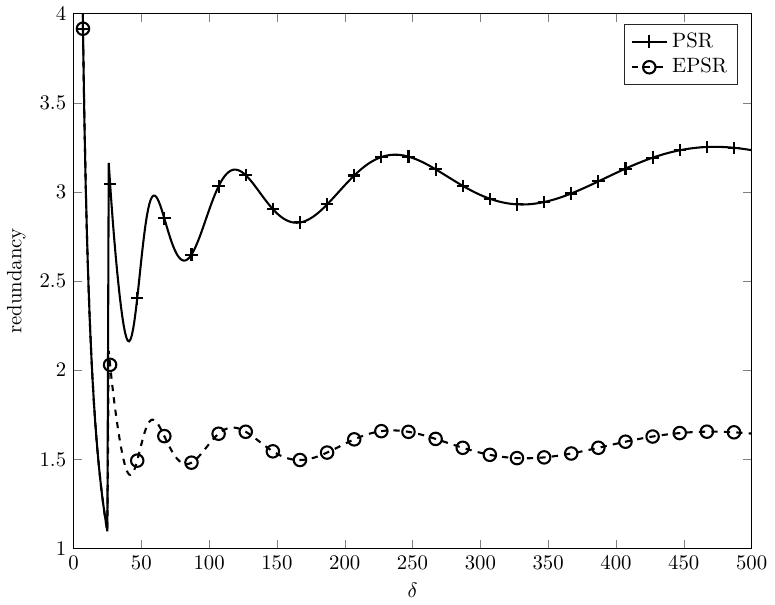}}  
        
    \subfloat[normalized complexity\label{fig:comp_vs_delta_comp}]{
        \includegraphics[width=0.99\columnwidth]{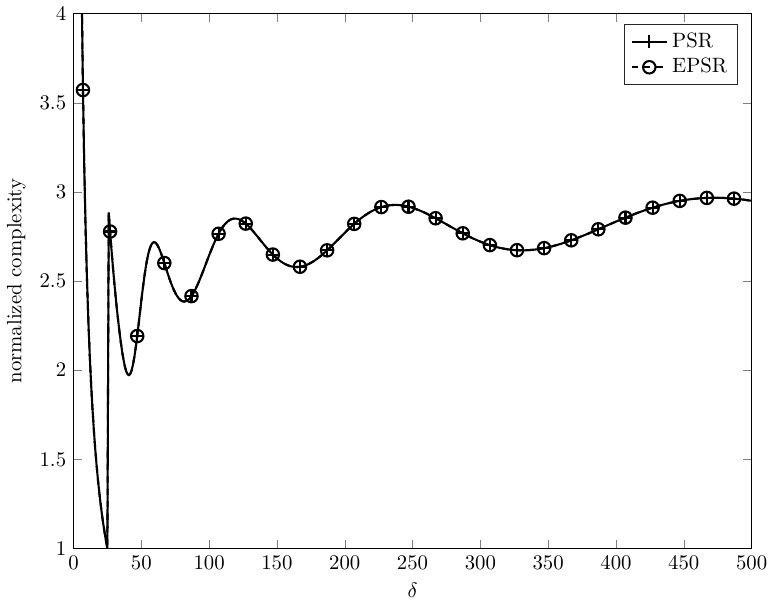}}
    \caption{Redundancy (a) and normalized complexity (b) as a function of $\d$ for \ac{PSR} and \ac{EPSR} with fair  partitioning. In all cases we assume $c=2$, $\bardelta=25$, $\kfact=1$ and $\ell=64$.}
    \label{fig:complexity_vs_delta_comparison}
\end{figure}

Fig.~\ref{fig:complexity_vs_delta_comparison} shows the redundancy (a) and normalized complexity (b) for \ac{PSR} and \ac{EPSR} as a function of the number of differences $\d$ for $\bardelta = 25$, $c=2$ and $p_1=p_2=1/2$. 
Fig.~\ref{fig:cost_vs_delta_comp} further demonstrates that the redundancy of both schemes oscillates around a specific value, indicating a linear relationship between communication cost and the set difference $\delta$.  The redundancy of \ac{EPSR} is approximately half that of \ac{PSR}. 
Specifically, \ac{PSR}'s redundancy oscillates around $\sim$3, while \ac{EPSR}'s oscillates around $\sim$1.5.\footnote{There is a direct analogy between the redundancy, as defined in this paper, and the inverse of the throughput (i.e., the normalized data delivery time) of tree algorithms in the context of random access, which is also demonstrated in the presented results for the redundancy. 
This is evident in the results presented. Specifically, the inverse throughput of binary tree algorithms without Successive Interference Cancellation (SIC) is approximately $2/\ln 2$, and with SIC it is about $1/\ln 2$. These values correspond respectively to redundancy values close to 3 and 1.5, as supported by \cite{Flajolet,yu2007high}.}
Thus, \ac{EPSR}'s redundancy represents a significant reduction compared to \ac{PSR}, much closer to the optimal value of 1.

Fig.~\ref{fig:comp_vs_delta_comp} shows the normalized complexity of both schemes.  Notably, the normalized complexities are identical. 
The approximately constant normalized complexity further suggests a linear relationship between time complexity and the number of differences $\d$.

We again note that for this comparison we have chosen the parameters $c=2$ and $p_1=p_2=1/2$, which, as discussed in Section~\ref{sec:param}, appear to be optimal for \ac{EPSR}, and also minimize redundancy in the case of \ac{PSR}, as shown in \cite{Minsky2002} and \cite{SGDK2020}. Therefore, this comparison between \ac{PSR} and \ac{EPSR} is arguably fair.

\subsection{Network Simulation}

In order to compare the \ac{PSR} and \ac{EPSR} algorithms in a realistic and controlled environment, a discrete-event simulation framework was developed in Python using the simpy library. The goal of this framework is evaluating the total time required for \ac{PSR} and \ac{EPSR} to reconcile two sets, in a network setting with a given latency, throughput and computational capabilities.

\begin{figure}[t]
    \centering    
    \includegraphics[width=0.99\columnwidth]{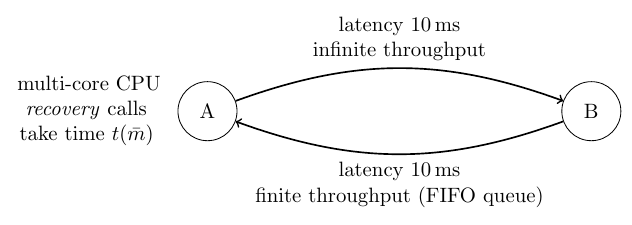}      
    \caption{Overview of the network simulation environment.}
    \label{fig:network}
\end{figure}

The simulation environment is depicted in Fig.~\ref{fig:network}.  We consider two network nodes, A and B. 
Nodes A and B are connected through a data network with a predefined one-way latency (e.g., 10~ms) that is the same in both directions. 
In the direction from A to B, the network is assumed to have infinite throughput, whereas in the direction from B to A, the network is assumed to have a predefined finite throughput\footnote{These assumptions are made for simplicity since node A only sends simple, short requests to node B, whereas node B replies to these requests by sending \ac{SR} data structures which can have a considerable size.}, and messages are sent through the network following their order of arrival, i.e., this link is modeled as a FIFO queue. 
In our simulation, we assume that the \ac{SR} data structures that have to be exchanged are precomputed in advance and are thus immediately available. 
Node A initiates the set reconciliation protocol, and is thus in charge of performing the computations associated with Algorithms~\ref{alg:psr} or \ref{alg:EPSR}.
In this respect, we assume that the computing complexity is dominated by the calls to the \rec operation.
We consider two different configurations regarding computing power. In the first one, we assume that recovery takes 12.3~ms, which is approximately the time recovery takes for $\bardelta=50$ using a modern laptop. Additionally, we also consider a low computing power scenario in which recovery takes 50 times longer, i.e., 615~ms.

Additionally, we assume that all other computing steps take zero time\footnote{This assumption is realistic for large values of $\bardelta$ since \rec is the only operation with complexity $\mathcal{O}(\bardelta^3)$, whereas all other operations have complexity $\mathcal{O}(\bardelta)$, see Table~\ref{tab:complexity}.}.
Further, node A is assumed to have a multi-core CPU with $n_{\text{core}}$ cores and it processes all computing tasks following the order in which they are initiated, i.e., the processors are modeled as a multi-server FIFO queue. In particular, the results are shown below for 1, 2, 4 and 8 cores.

We consider three different network scenarios. In all three cases, we considered a network latency of 10 ms.
Scenario~I is characterized by high throughput (100 Mbps) and  high computing power (\rec takes 12.3~ms), 
scenario~II exhibits high throughput (100 Mbps) and low computing power (\rec takes 615~ms) and finally, scenario~III is characterized by low throughput (10 kbps) and high computing power (\rec takes 12.3~ms).
Table~\ref{tab:network} shows the parameters of the different scenarios considered. 
In all cases, we considered the set elements to be random 32-byte strips (256 bits), and we used $\bardelta=50$,  $\kfact=1$, and binary fair partitioning, i.e., $c=2$ and $p_1=p_2=1/2$.
For each scenario and value of the set difference $\d$, we generated 100 different samples (random partitions of the set difference), and thus obtained an estimation of the average time needed to reconcile all set differences.

\begin{figure}[!ht]
    \centering
    \subfloat[scenario I\label{fig:scenario_I}]{
        \includegraphics[width=0.97\columnwidth]{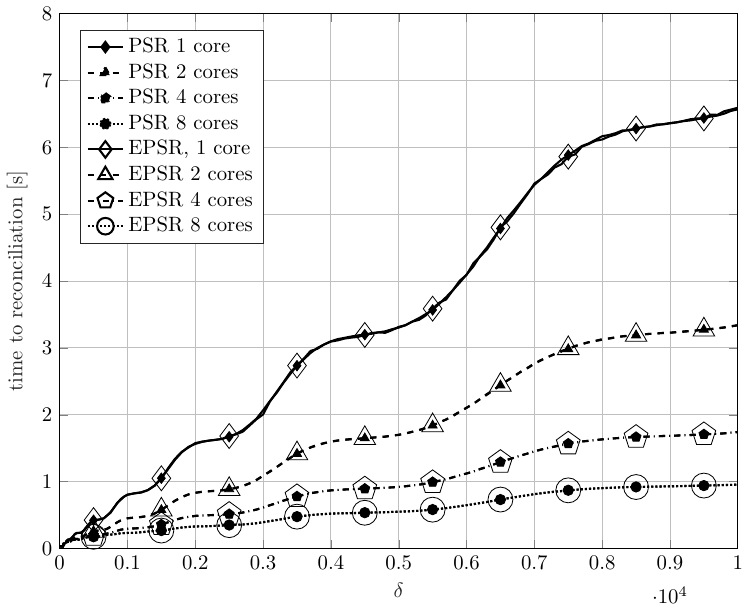}}  
        
    \subfloat[scenario II\label{fig:scenario_II}]{
        \includegraphics[width=0.97\columnwidth]{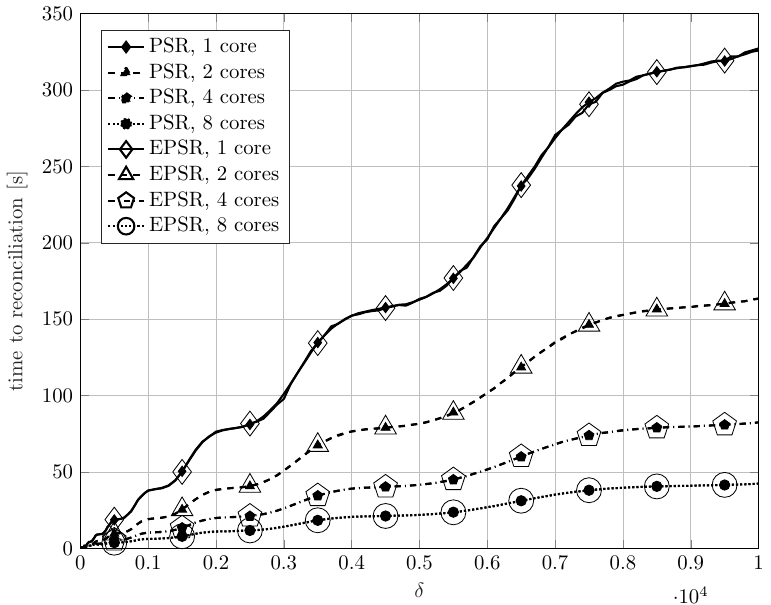}}  
    
    \subfloat[scenario III\label{fig:scenario_III}]{
        \includegraphics[width=0.97\columnwidth]{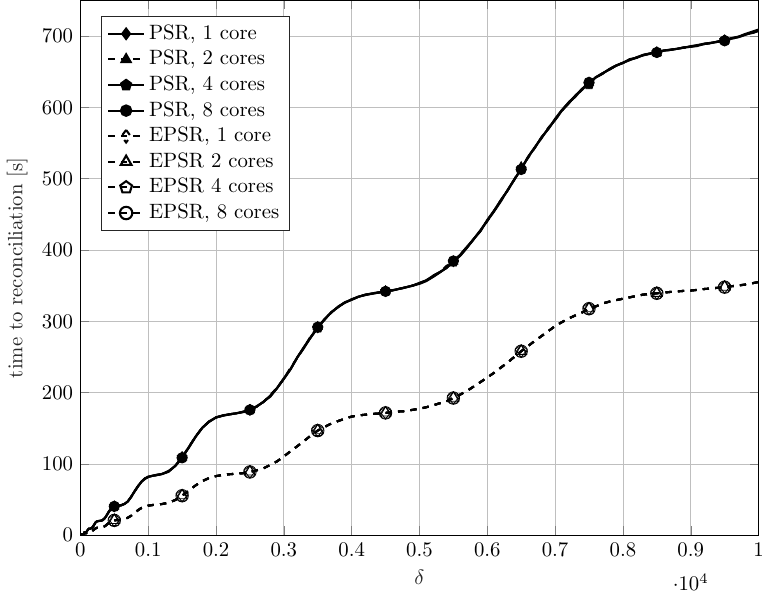}}  
        
    \caption{Total time to reconciliation as a function of the number of differences $\d$ for the 3 different scenarios.}
    \label{fig:scenarios}
\end{figure}

\begin{table}[t]
\centering
\begin{tabular}{lccc}
\toprule
parameter & scenario I & scenario II & scenario III\\
\midrule
latency [ms] & 10 & 10 & 10 \\
throughput [bps] & $10^{8}$ & $10^{8}$ & $10^{4}$ \\
recovery runtime [ms] & 12.3 & 615 & 12.3 \\
\bottomrule
\end{tabular}
\caption{Network configurations considered}
\label{tab:network}
\end{table}

Fig.~\ref{fig:scenarios} shows the total time to reconciliation for the 3 different network scenarios considered. 
Fig.~\ref{fig:scenario_I} shows the results for scenario~I, i.e., high-throughput and high-computing-power. As can be observed for this scenario, if the number of computing cores is kept constant, the total time to reconciliation of \ac{PSR}  and \ac{EPSR} are the same.  
This is explained by the fact that this scenario is latency limited:  the round-trip time is 20 ms, whereas the time needed for one call to recovery is 12.3 ms, and the time needed to transmit the \ac{SR} data structure is 0.13 ms. 
Since the number of communication rounds is the same for both protocols, we obtain the same latency. The results also indicate that, in this scenario, increasing the number of computing cores decreases the total time to reconciliation. In particular, doubling the number of cores almost halves the total time to reconciliation. 

Fig.~\ref{fig:scenario_II} shows the results for scenario~II, i.e., high-throughput and low-computing-power. As can be observed, for this scenario, if we keep the number of computing cores constant, the total time to reconciliation of \ac{PSR}  and \ac{EPSR} are again basically the same. 
In this case, the limiting factor is the execution time of the \rec call, which takes 615 ms and dominates over the round-trip time of 20 ms and the transmission time of 0.13 ms. This result is in line with Fig.~\ref{fig:complexity_vs_delta_comparison} that shows that \ac{PSR} and \ac{EPSR} have the same normalized complexity, i.e., they call recovery the same number of times. If we compare the curves for different numbers of computing cores, we can see again how doubling the number of computing cores almost halves the total time to reconciliation. 

Finally, Fig.~\ref{fig:scenario_III} shows the results for scenario~III, i.e., low-throughput and high-computing-power. For this scenario, if we keep the number of computing cores constant, the total time to reconciliation of \ac{PSR} almost doubles that of \ac{EPSR}.  In this case, the transmission time of the \ac{SR} data structure is 1.3~s, and this dominates over the round-trip time of 20\,ms and the 12.3\,ms computing time associated with a \rec call. These results are in line with Fig.~\ref{fig:cost_vs_delta_comp}, which indicates that the communication cost of \ac{PSR} almost doubles that of \ac{EPSR}. Interestingly, in this bandwidth-limited regime, increasing the number of computing cores does not reduce the total time to reconciliation.

In summary, the results of our network simulations indicate that when the bottleneck resides in the latency or computation time, binary \ac{PSR} and {EPSR} result in the same time to reconciliation. Additionally, in such scenarios, having a  CPU with multiple computing cores decreases the total time to reconciliation. 
In contrast, when the bottleneck resides in the communication bandwidth between the nodes, the time to reconciliation of \ac{EPSR} is lower than that of \ac{PSR}. In addition, having multiple computing cores does not reduce the total time to reconciliation in this scenario. 
All these results are in line with our analysis in Section~\ref{sec:analysis}. 

Finally, we note that in our network simulation setup, only two nodes have been considered for simplicity, while in many practical set reconciliation setups, there will be many clients (like Node A) that would send requests to a server (Node B) to synchronize their sets with that of the server. In such a setup, since the processing takes place at the clients, the limiting factor in the performance of the set reconciliation algorithm could very well be the server bandwidth.
The presented results indicate that the use of \ac{EPSR} could be advantageous in such scenarios, however, detailed simulations of such a setup are beyond the scope of this work.

\section{Conclusions}\label{sec:conclusions}

In this paper, we have proposed a novel set reconciliation algorithm termed \acf{EPSR}. As its name indicates, this algorithm is an improvement of the well-known \acf{PSR} algorithm, that relies on a divide-and-conquer approach in which the sets to be reconciled are successively partitioned until they
contain a number of differences below some predetermined value.
Our analysis shows that \ac{EPSR} exhibits the same time and communication round complexity as \ac{PSR} (which is linear in the number of differences) while reducing  the communication cost by almost a factor of 2.
Additionally, we also evaluated the performance of \ac{EPSR} in a simulated network environment and compared it to \ac{PSR} in terms of the total time needed for the protocols to synchronize two sets. The simulations showed that \ac{EPSR} outperforms \ac{PSR} when the limiting factor is the communication bandwidth. 
We believe that this novel algorithm can find applications in areas such as remote file synchronization protocols \cite{agarwal2006bandwidth} and cryptocurrency networks \cite{Ozisik2019, bovskov2022gensync}, particularly in scenarios with limited communication bandwidth, where minimizing communication costs is crucial.

\FloatBarrier

\appendices

\section{Proof of Theorem~\ref{thm:PSR}}\label{app:PSR}

Recall that $\npc_\d $ denotes the random variable associated with the number of calls to the procedure \rec when executing \ac{PSR}. 
We can write
\begin{align}
\label{eq:rec_psr}
    \npc_\d = \begin{cases}
        1 & 0 \leq \d \leq \bardelta \\
        1 + \sum_{j = 1}^{c} \npc_{i_j} & \text{else}
    \end{cases} 
\end{align}
where $0 \leq i_j \leq \d$, $j = 1, 2, \dots c$, is the (random) number of differences out of $\d$ that ended in $j$-th partition, $\sum_{j = 1}^{c}i_j = \delta$, and $\npc_{i_j}$ is the number of \rec calls for the $j$-th partition.

For the ease of exposition, we introduce the following notation
\begin{align}
    & \mathbf{i}_k  = [ i_1, \dots, i_k ] , \; i_j \geq 0, j = 1, \dots, k \\
    & \mathbf{i}_{k,l}  = [ i_1, \dots, i_{l-1}, i_{l+1}, \dots i_k ], \; l=1, \dots, k\\
    & \binom{ d} {\mathbf{i}_k }  = \binom{d}{i_1, \dots, i_k} \\  
    & \mathcal{A}_k^d  = \{ \mathbf{i}_k | \sum_{j = 1}^{k} i_j = d \} \\
    & \mathcal{A}_{k,l}^d = \{ \mathbf{i}_{k} | \sum_{j = 1, j \neq l}^{k} i_j = d \} 
\end{align}
We continue by leveraging the approach presented in~\cite{Flajolet} and subsequently expanded in~\cite{yu2007high,SGDK2020}.
First, we compute the conditional \ac{PGF} of $\npc_\d$, denoted by $\mypgf{\npc_\d}(z) = \mathrm{E} \left\{ z^{\npc_\d} \right\}$.
It is easy to verify that $\mypgf{\npc_\d} (z) = z$ for $0 \leq \d \leq \bardelta$.
Further, for $\d > \bardelta$ 
\begin{align}
\mypgf{\npc_\d}(z) & = \mathrm{E}  \left[ z^{1 + \sum_{j=1}^{c} \npc_{i_j} } \Big| \mathbf{i}_c \in \mathcal{A}_c^\d  \right]   = z \, \mathrm{E} \left[  \prod_{j=1}^{c} z^{\npc_{i_j}} \Big| \mathbf{i}_c \in \mathcal{A}_c^\d \right]  \\
& = z \sum_{ \mathbf{i}_c \in \mathcal{A}_c^\d } \binom{\d}{\mathbf{i}_c } \prod_{k=1}^{c} p_k^{i_k} \mypgf{\npc_{i_k}}(z).
\end{align}

The conditional expectation $\expnpc_\d$ is obtained by taking the first derivative of $\mypgf{\npc_\d}(z)$ and equating it with 1
\begin{align}
    \expnpc_\d  = & \frac{ \partial \mypgf{\npc_\d}(z) }{\partial z} \Big|_{z=1} \\
     = & \sum_{ \mathbf{i}_c \in \mathcal{A}_c^\d } \binom{\d}{\mathbf{i}_c } \left\{ \prod_{k=1}^{c} p_k^{i_k}   + \sum_{k=1}^{c} \frac{\partial \mypgf{\npc_{i_k}}(z)} {\partial z} \Big|_{z=1} \prod_{l=1}^{c} p_l^{i_l} \right\} \\
     = & 1 + \sum_{k=1}^{c} \sum_{i_k = 0}^{\d} \binom{\delta}{i_k} p_k^{i_k} \expnpc_{i_k} \sum_{\mathbf{i}_{c,k} \in \mathcal{A}_{c,k}^{\d - i_k } } \binom{ \d - i_k} {\mathbf{i}_{c,k} }  \prod_{\substack{l=1 \\ l \neq k}}^{c} p_l^{i_l} \\
     = & 1 + \sum_{k=1}^{c} \sum_{i_k=0}^{\d} \binom{\d}{i_k} p_{k}^{i_k}( 1 - p_{i_k})^{\d - i_k} N_{i_k} \label{eq:exnpc_t}
\end{align}
where we used the fact that $\expnpc_{i_k} = \frac{\partial \mypgf{\npc_{i_k}}(z)} {\partial z} \Big|_{z=1}$.
Rearranging Eq.~\eqref{eq:exnpc_t} and substituting 
$k$ by $j$, and $i_k$ by $i$, we obtain
\begin{align}
    \expnpc_\d = \frac{ 1 + \sum_{j=1}^{c} \sum_{i = 0}^{\d - 1} {\d \choose i } p_j^{i} ( 1 - p_j)^{\d - i_j} \expnpc_{i} }{ 1 - \sum_{j=1}^{c} p_j ^\d}
\end{align}
which verifies Eq.~\eqref{eq:expnpc}.

\hfill $\square$


\section{Proof of Lemma~\ref{lem:psr_round}}
\label{app:psr_round}
The number of communication rounds of \ac{PSR} corresponds to the depth of the partition tree. Let us denote by $\Depth$ the random variable associated with the partition tree.
Let us observe that a node at depth $\depth$ can be identified by means of a length-$\depth$ word  $\bm{w}= (w_1, \dots, w_\depth)$, with $w_i \in \{1,\dots c\}$.
For example, consider a node associated with a word $\bm{w}= (1, 3, 2)$. This indicates that the corresponding set element is at depth 3. Additionally, the element ended up in partition 1 after the first set partition. After the second partition, the element resided in partition 3, and after the third and final partition, the element is in partition 2.
The depth $\Depth$ can be formally defined as follows,
\[
\Depth \coloneq \min \{ \depth \geq 0 | \max_{|\bm{w}|=d} S_{\bm{w}}^{(d)}\leq \bardelta\}
\]
where $S_{\bm{w}}^{(d)}$ is the number of elements of the set difference that are in the partition associated with the word $\bm{w}$. We have that $S_{\bm{w}}^{(d)} \sim  \mathrm{Binomial}(\d, q_{\bm{w}})$, where  $q_{\bm{w}}\coloneq \prod_{k=1}^\depth p_{w_k}$ is the probability associated with the path leading to the node whose associated word is $\bm{w}$.

Let us define $\pmax \coloneq \max_{i=1\dots c} p_i$ and ${\lambda \coloneq - \frac{1}{\log (\pmax)}}$. The \emph{heaviest} node of the tree (partition) at depth $\depth$ has probability $\pmax^\depth$, and its \emph{load} (expected number of elements in that partition) is $\mu_d =\mathrm{E} [S_{\bm{w}}^{(d)}] = \d \pmax^\depth$.
Define $\beta_{\depth} \coloneq \bardelta/ \mu_d$ and ${h(\beta_{\depth}) \coloneq \beta_{\depth} \log \beta_{\depth} +1}$.
The probability of the heaviest node of the tree at depth $\depth$ having more than $\bardelta$ elements can be upper bounded by $e^{- \mu_d h(\beta_{\depth})}$ (this is the Chernoff bound).
If we now take a union bound over all the $c^\depth$ partitions at depth $\depth$, we obtain the following bound on the depth of the tree
\begin{equation}
  \Pr (\Depth > \depth) = \Pr (S_{\bm{w}}^{(d)} > \bardelta ) \leq c^\depth e^{- \mu_d h(\beta_{\depth})}.  
  \label{eq:chernoff}
\end{equation}
We can now apply the sum-tail formula to rewrite $ \mathrm{E} [\Depth]$ as
\begin{equation}    
\mathrm{E} [\Depth] = \sum_{d=0}^\infty \Pr (\Depth > \depth) 
\label{eq:tail_sum}
\end{equation}

Define now $\depth_0$ as the first level at which we expect the load of the heaviest partition to drop below $\bardelta$,  $\depth_0 \coloneq \lceil \lambda  \log \frac{\d}{\bardelta} \rceil$.
We can break the summation in Eq.~\eqref{eq:tail_sum} as follows
\begin{equation}    
\mathrm{E} [\Depth] = \sum_{d=0}^{\depth_0-1} \Pr (\Depth > \depth) + \underbrace{\sum_{d=\depth_0}^{\infty} \Pr (\Depth > \depth) }_{\tau}
\label{eq:depth_0}
\end{equation}    

The first term in Eq.~\eqref{eq:depth_0} can be upper bounded as 
\begin{equation}
\sum_{d=0}^{\depth_0-1} \Pr (\Depth > \depth) \leq \sum_{d=0}^{\depth_0-1} 1 = \depth_0 = \lceil \lambda  \log \frac{\d}{\bardelta} \rceil
\label{eq:depth_1}.
\end{equation}

Let us now look into the second term in Eq.~\eqref{eq:depth_0} (the tail). 
Consider a tree depth  $\depth= \depth_0 + k$, with $k \geq 0$.
We have that $\mu_\depth = \mu_{\depth_0} \pmax ^k \leq \bardelta \pmax^k$. Similarly, $\beta_{\depth} = \bardelta / \mu_{\depth} \geq \pmax^k$.
If $\beta_\depth \geq e$, we lower bound $h(\beta_\depth)$ as
\begin{equation}
    h(\beta_\depth) \geq \frac{1}{2} \beta_\depth \log \beta_\depth.
    \label{eq:beta_bound}
\end{equation}
Let us now define as $k^\star$ the value of $k$ for which $\beta_\depth \geq e$. Observe that, by definition, $\pmax \geq 1/c$. Hence, we have $k^\star=2$ for $c=2$ and $k^\star=1$ for $c>2$. We can rewrite the tail in Eq.~\eqref{eq:depth_1} as
\begin{align}
  \tau &=   \sum_{d=\depth_0}^{\infty} \Pr (\Depth > \depth) \\
  &= \underbrace{\sum_{k =0}^{k^\star -1} \Pr (\Depth > \depth_0 +k)}_{\tau^\prime} + 
  \underbrace{\sum_{k =k^\star}^{\infty} \Pr (\Depth > \depth_0 +k)}_{\tau^{\prime \prime}}.
  \label{eq:tau_tau_prime}
\end{align}
We have that $\tau^\prime \leq k\star$, and thus this first term is $\mathcal{O}(1)$.
Making use of our bounds in Eqs.~\eqref{eq:chernoff} and \eqref{eq:beta_bound} the second term can be upper bounded as
\[
\tau^{\prime \prime} \leq c^{\depth_0} e^{- \frac{1}{2} \mu_{\depth} \beta_\depth \log \beta_\depth}
\]
Relying on $\beta_k \geq \pmax^k$ we can upper bound $\tau^{\prime \prime}$ as
\[
\tau^{\prime \prime} \leq c^{\depth_0} c^k e^{- \frac{1}{2} \bardelta k \log\frac{1}{\pmax}}
\]
which can be rewritten as 
\[
\tau^{\prime \prime} \leq c^{\depth_0} e^{- (1-\theta) k \log\frac{1}{\pmax}}
\]
where $\theta \coloneq \frac{\log c}{\frac{1}{2} \bardelta \log\frac{1}{\pmax}}$.
Now, we can distinguish two cases based on $\theta$. If $\theta<1$ (which implies ${\bardelta> \frac{2 \log c}{\log\frac{1}{\pmax}}}$), we have that the tail $\tau^{\prime \prime}$ is a geometric series that sums up to $\mathcal{O}(1)$.
Alternatively, if $\theta>1$ we have that the sum diverges. However, in that case, we can again split the sum in the expression of  $\tau^{\prime \prime}$.

\begin{align}
  \tau^{\prime \prime} &=     &= \underbrace{\sum_{k =k^\star}^{k^\dagger-1} \Pr (\Depth > \depth_0 +k)}_{\tau^\dagger} + 
  \underbrace{\sum_{k =k^\dagger}^{\infty} \Pr (\Depth > \depth_0 +k)}_{\tau^{\ddagger}}.
  \label{eq:tau_prime_prime}
\end{align}
where $k^\dagger \coloneq \lceil \frac{2 \log c}{\log (1/\pmax)} \log \log (\d/\bardelta)\rceil$.
Using the same arguments we used before, we have that $\tau^\dagger$ is upper bounded by $k^\dagger$ and hence it is $\mathcal{O}(\log\log \d/\bardelta)$. Additionally, we can now plug in our Chernoff bound in \eqref{eq:chernoff} in the expression of $\tau^{\ddagger}$. This yields a geometrically decaying sequence, whose sum is $\mathcal{O}(\log^{-2} (\delta/\bardelta))$.
Putting it all together we have that 
\[
\mathrm{E} [\Depth] = \mathcal{O} \left( \lambda  \log \frac{\d}{\bardelta}\right) + \mathcal{O} (1) + \mathcal{O}(\log\log \d/\bardelta)
\]
where the dominant term is $\mathcal{O} \left( \lambda  \log \frac{\d}{\bardelta}\right)$, which finalizes the proof.


\section {Proof of Theorem~\ref{thm:EPSR}}
\label{app:EPSR}

The example presented in Fig.~\ref{fig:tree_sic_m_2} shows that, in the \ac{EPSR} algorithm with $c=2$, the data structures associated with right child nodes are not sent over the network.
In the general case where $c \geq 2$, assuming that the parent partition contains $\d$ differences, \ac{SR} data structures will only be sent for the children nodes $j = 1,\dots,\epsrskip$,  where
\begin{align}
\label{eq:ell}
\epsrskip = \min \left\{ k \in \{1,\dots,c\} \Big| \sum_{j=1}^{k} i_j \geq \d - \bardelta \right\}.
\end{align}
In other words, as soon as the accumulated number of reconciled differences in the child nodes, starting from the leftmost child node (partition), becomes at least $\d - \bardelta$, the remaining differences will be reconciled in the parent node and the remaining child node will be skipped.
We also note that the rightmost child node, $\epsrskip =c$, is always skipped\footnote{When the last child node (i.e.,  $\epsrskip =c$) is reached, we have that ${\sum_{j=1}^{c-1} i_j < \d - \bardelta}$, so there are more than $\bardelta$ differences left in the partition associated with this last child node. Thus, the \ac{SR} data structure associated with this node (partition) is not sent over the network, since we already have it.},
and the  partition associated with the node can be further partitioned.

The recursive relation for the number of \ac{SR} data structures sent over the network is now
\begin{align}
\label{eq:npc_sic}
    \npce_\d = \begin{cases}
        1 & 0 \leq \d \leq \bardelta \\
        \mathbf{1}\{ \epsrskip < c \} + \sum_{j = 1}^{\epsrskip} \npce_{i_j} & \text{else}
    \end{cases} 
\end{align}
where $\mathbf{1}\{ \epsrskip < c \}$ is an indicator function whose value is 1 if ${\epsrskip < c}$ and 0 otherwise.
The use of the indicator function reflects the fact that if $\epsrskip = c$, then the last child node is skipped, and the skipped \ac{SR} data structure call should be subtracted from the sum in Eq.~\eqref{eq:npc_sic}.

The rest of the proof roughly follows the approach described in~\cite{VDSK2024}.\footnote{The approach in~\cite{VDSK2024} was used to derive the \emph{direct} formula for $\expnpce_\d$ for the case of \emph{binary} partition.}
For ease of exposition, we define $F_k = \sum_{l = 1}^{k} p_l$.

Denote the \ac{PGF} of $\npce_\d$ by $\mypgf{\npce_\d} (z)$. 
It is easy to verify that $\mypgf{\npce_\d} (z) = z$ for $0 \leq \d \leq \bardelta$.
For $\d > \bardelta$, we can write
\begin{align}
    \mypgf{\npce_\d}(z) & = \mathrm{E}  \left[ z^{\npce_\d} \right] =  \sum_{k=1}^{c} \mathrm{E}  \left[ z^{\npce_\delta} , \epsrskip = k \right]
\end{align}
where $\npce_\d$ is given by Eq.~\eqref{eq:npc_sic}.

For $\epsrskip=1$, it can be shown that
\begin{align}
    \mathrm{E}  & \left[ z^{\npce_\delta} , \epsrskip = 1 \right] = \sum_{j=0}^{\bardelta} \mathrm{E}  \left[ z^{\npce_\delta} , \epsrskip = 1, i_1 = \delta - j \right]  \label{eq:1}\\
    & = z \sum_{j=0}^\bardelta \binom{\d}{j} (1 - F_1)^j \sum_{\mathbf{i}_1 \in \mathcal{A}_1^{(\d - j)}} \binom{\d - j}{\mathbf{i}_1} p_1^{i_1} \mypgf{\npce_{i_1}}(z). 
\end{align}
The equation reflects the fact that, when $\epsrskip = 1$, no more than $\bardelta$ differences can end up in children other than the first one.

For $\epsrskip=k$, $k = 2, \dots, c -1$, it can be shown that
\begin{align}
    \mathrm{E}  & \left[ z^{\npce_\delta} , \epsrskip = k \right] = \sum_{j=0}^{\bardelta} \mathrm{E}  \left[ z^{\npce_\delta} , \epsrskip = k, \sum_{l=1}^{k} i_l = \delta - j \right ] \label{eq:k} \\
    & = z \sum_{j=0}^\bardelta \binom{\d}{j} (1 - F_k)^j \sum_{\mathbf{i}_k \in \mathcal{A}_k^{(\d - j)}} \binom{\d - j}{\mathbf{i}_k} \prod_{l=1}^{k} p_l^{i_l} \mypgf{\npce_{i_l}}(z) \, - \\
    & \phantom{=} \, \, \, z^2 \sum_{j=0}^\bardelta \binom{\d}{j} (1 - F_{k-1})^j \mkern-20mu \sum_{\mathbf{i}_{k-1} \in \mathcal{A}_{k-1}^{(\d - j)}} \mkern-5mu \binom{\d - j}{\mathbf{i}_{k-1}} \prod_{l=1}^{k - 1} p_l^{i_l} \mypgf{\npce_{i_l}}(z). 
\end{align}
The first term in Eq.~\eqref{eq:k} corresponds to the partition configurations in which the $\d - j$ differences ended up in the first $k$ children.
The second term corresponds to the subset of these configurations in which there are $\d - j$ differences in the first $k - 1$ children (note that here we have $\mypgf{\npce_{i_k}}(z) = z$), which are then subtracted from the first term.

Finally, for $\epsrskip = c$, we have
\begin{align}
    \mathrm{E}  & \left[ z^{\npce_\delta} , \epsrskip = c \right] = \sum_{\mathbf{i}_c \in \mathcal{A}_c^{(\d)}} \binom{\d}{\mathbf{i}_c} \prod_{l=1}^{c} p_l^{i_l} \mypgf{\npce_{i_l}}(z) \, - \label{eq:c}\\
    & z \sum_{j=0}^\bardelta \binom{\d}{j} (1 - F_{c-1})^j \sum_{\mathbf{i}_{c-1} \in \mathcal{A}_{c-1}^{(\d - j)}} \binom{\d - j}{\mathbf{i}_{c-1}} \prod_{l=1}^{c - 1} p_l^{i_l} \mypgf{\npce_{i_l}}(z) 
\end{align}
as in this case it has to be that $\sum_{l=1}^{c} i_l = \delta$.

Combining Eqs.~\eqref{eq:1},\eqref{eq:k}, and \eqref{eq:c}, we get
\begin{align}
    & \mypgf{\npce_\d}(z) =  \sum_{\mathbf{i}_c \in \mathcal{A}_c^{(\d)}} \binom{\d}{\mathbf{i}_c} \prod_{l=1}^{c} p_l^{i_l} \mypgf{\npce_{i_l}}(z) \, + ( z - z^2) \times \\
    & \qquad \qquad \sum_{k=1}^{c-2} \sum_{j=0}^\bardelta \binom{\d}{j} (1 - F_{k})^j 
    \mkern-20mu \sum_{\mathbf{i}_{k} \in \mathcal{A}_{k}^{(\d - j)}} \mkern-5mu \binom{\d - j}{\mathbf{i}_{k}} \prod_{l=1}^{k} p_l^{i_l} \mypgf{\npce_{i_l}} (z)
\end{align}
from which we have 
\begin{align}
    \expnpce_\d  & = \frac{ \partial \mypgf{\npce_\d}(z) }{\partial z} \Big|_{z=1} \mkern-2mu \\
     & = \mkern-4mu \sum_{k=1}^c \sum_{i_k = 0}^\d \binom{\d}{i_k} p_k^{i_k} ( 1 - p)k)^{\d - i_k} \expnpce_{i_k} - \\
    & \phantom{=} \sum_{k=1}^{c-2} \sum_{j=0}^\bardelta \binom{\d}{j} (1 - F_{k})^j  \mkern-20mu \sum_{\mathbf{i}_{k} \in \mathcal{A}_{k}^{(\d - j)}} \mkern-10mu \binom{\d - j}{\mathbf{i}_{k}} \prod_{l=1}^{k} p_l^{i_l} \\
    & = \sum_{k=1}^c \sum_{i_k = 0}^\d \binom{\d}{i_k} p_k^{i_k} ( 1 - p_k)^{\d - i_k} \expnpce_{i_k} - \\
    & \phantom{=} \, \, \sum_{k=1}^{c-2} \sum_{j=0}^\bardelta \binom{\d}{j} (1 - F_{k})^j F_k^{\d - j}.
\end{align}
Rearranging, and substituting $k$ by $j$, and $i_k$ by $i$, we get
\begin{align}
    \expnpce_\d  = &\frac{\sum_{j=1}^{c} \sum_{i = 0}^{\d - 1} {\d \choose i } p_j ^{i} ( 1 - p_j )^{\d - i } \expnpce_{i}  }{ 1 - \sum_{j=1}^c p_j^\d}   \\
    - &  \frac{ \sum_{j=1}^{c-2} \sum_{i=0}^\bardelta \binom{\d}{i} (1 - F_j)^i F_j^{\d - i}}{1 - \sum_{j=1}^c p_j^\d}.
\end{align}

\hfill $\square$

\section{Proof of Theorem~\ref{thm:EPSRrec}}
\label{app:EPSR_rec}

The proof of Theorem~\ref{thm:EPSRrec} goes along the lines similar to the proof of Theorem~~\ref{thm:EPSR}.

Let us denote by $\nrec_\d$ the random variable associated with the number of calls to the \rec procedure in \ac{EPSR} when the number of differences between the two sets is $\d$.
If $\d \leq \bardelta$ we obviously only need one call to \rec. For $\d > \bardelta$, we perform the following analysis. 
Let $i_j$, with $0 \leq i_j \leq \d$, denote the number of differences from the initial $\d$ that ended up in the $j$-th partition. Hence, we have $\sum_{j = 1}^{c}i_j = \delta$. 
The first recovery call happens in line 4 of Algorithm~\ref{alg:EPSR_r}, when \rec is called.
If $\epsrskip < c$, there will also be $\sum_{j = 1}^{\epsrskip} \nrec_{i_j}$ calls to \rec through recursive invocation of EPSR in line 11, as well as $\epsrskip$ \rec calls in line 15 of Algorithm~\ref{alg:EPSR_r}.
If $\epsrskip = c$, then there will also be $\sum_{j = 1}^{\epsrskip} \nrec_{i_j} - 1$  calls to \rec, as the call of EPSR in line 20 for the last $c$-th partition is with $\flagskip$ flag set to true, and, similarly, there will be only additional $\epsrskip - 1$ calls of \rec, as the procedure is not called for the last, $c$-th partition.  
Hence, we can write
\begin{align}
\label{eq:nrec_sic}
    \nrec_\d = \begin{cases}
        1 & 0 \leq \d \leq \bardelta \\
        \mathbf{1}\{ \epsrskip < c \} + \sum_{j = 1}^{\epsrskip} \nrec_{i_j} + \epsrskip - \mathbf{1}\{ \epsrskip = c \}& \text{else}
    \end{cases} 
\end{align}

Obviously, the first two terms in Eq.~\eqref{eq:nrec_sic}, i.e, $\mathbf{1}\{ \epsrskip < c \} + \sum_{j = 1}^{\epsrskip} \nrec_{i_j}$,  are functionally identical to the ones in Eq.~\eqref{eq:npc_sic}.
Thus, their contribution to the expected number of recovery calls, denoted by $\expnrec_\d  =  \mathrm{E} \left[  \nrec_\d \right]$, will be 
\begin{align}
\label{eq:first_two}
    & \sum_{k=1}^c \sum_{i_k = 0}^\d \binom{\d}{i_k} p_k^{i_k} ( 1 - p_k)^{\d - i_k} \expnrec_{i_k} - \\
    & \phantom{=} \, \, \sum_{k=1}^{c-2} \sum_{j=0}^\bardelta \binom{\d}{j} (1 - F_{k})^j F_k^{\d - j}
\end{align}

The expectation of the third and fourth term of Eq.~\eqref{eq:nrec_sic},  i.e., $\epsrskip - \mathbf{1}\{ \epsrskip = c \}$,  is as follows
\begin{align}
& \mathrm{E} \left[ h | \delta \right] - \mathrm{P} [ h = c | \d ] = \sum_{j=1}^{c} j \, \mathrm{P} [ \epsrskip = j | \d ] - \mathrm{P} [ h = c | \d ] \\
& \,= c - \sum_{k=1}^{c-1} \sum_{j=1}^k \mathrm{P} [ \epsrskip = j | \d ]   -  1 + \sum_{j }^{c-1} \mathrm{P} [ h = j | \d ] \\
& \,=  c - 1 - \sum_{k=1}^{c-2} \sum_{j = 0}^{\bardelta} \binom{\d}{j} (1 - F_{k})^j F_k^{\d - j}. \label{eq:second_two}
\end{align}

Combining Eqs.~\eqref{eq:first_two} and \eqref{eq:second_two}, we get
\begin{align}
    \expnrec_\d = & c  - 1 
     - 2 \sum_{k=1}^{c-2} \sum_{j = 0}^{\bardelta} \binom{\d}{j} (1 - F_{k})^j F_k^{\d - j } \\
    & + \sum_{k=1}^c \sum_{i_k = 0}^\d \binom{\d}{i_k} p_k^{i_k} ( 1 - p_k)^{\d - i_k} \expnrec_{i_k}
\end{align}
which, after some manipulation, yields
\begin{align}
    \expnrec_\d  = &\frac{\sum_{j=1}^{c} \sum_{i = 0}^{\d - 1} {\d \choose i } p_j ^{i} ( 1 - p_j )^{\d - i } \expnrec_{i}  }{ 1 - \sum_{j=1}^c p_j^\d}   \\
    & +   \frac{ c - 1 - 2 \sum_{j=1}^{c-2} \sum_{i=0}^\bardelta \binom{\d}{i} (1 - F_j)^i F_j^{\d - i} }{1 - \sum_{j=1}^c p_j^\d}. 
\end{align}
\hfill $\square$

\section{Proof of Lemma~\ref{lemma:epsr_rounds} }
\label{app:EPSR_rounds}
\emph{Intuition.} We map the $c$-ary EPSR step into a sequence of binary splits.  First $P$ is split into $\big(P_1,\ \bigcup_{i>1} P_i\big)$.  If the right branch fails, it is split into $\big(P_2,\ \bigcup_{i>2} P_i\big)$, and so on. This yields $c-1$ binary splits with left-branch probability ${p_i^\star = \frac{p_i}{1-\sum_{j<i} p_j} }$. The number of rounds is upper-bounded by the worst split along this chain, and its recursion depth is then bounded using Lemma~\ref{lem:psr_round}.

Consider \ac{EPSR} with partitioning factor $c>2$, and assume that recovery fails for partition $\P$.
Thus, we partition $\P$ into $c$ partitions $\P_i, i=1,\dots, c$, where $p_i$ corresponds to the probability that a randomly chosen element in $\P$ is at partition $i$ after partitioning. 
According to Algorithm~\ref{alg:EPSR_r}, EPSR will first process 2 partitions in parallel, namely $\P_1$ and $\cup_{i>1} \P_i = \P \setminus \P_1$. 
Each of the elements in $\D \cap \P$ will be in $\P_1$ with probability $p_1$ and with prob $1-p_1$ in $\cup_{i>1} \P_i$.
If recovery fails in $\cup_{i>1} \P_i$, the algorithm will process in parallel two partitions:  $\P_2$ and $\cup_{i>2} \P_i$. The elements in  $\D \cap \left( \cup_{i>1} \P_i \right)$ will be in $\P_2$ with probability $p_2^\star = \frac{p_2}{1- p_1}$, and with probability $1-p_2^\star$ they will be in partition $\cup_{i>2} \P_i$. 
In case recovery fails in partition $\cup_{i>2} \P_i$, the algorithm will process in parallel two partitions: $\P_3$ and $\P_4$. Elements in  $\D \cap \left( \cup_{i>2} \P_i \right)$ will be in $\P_3$ with probability $p_3^\star = \frac{p_3}{1- p_1 + p_2}$,  and with probability $1-p_3^\star$ they will be in partition $\P_4$. 

\begin{figure}[t]
	\centering
	\scalebox{0.73}{
		\begin{tikzpicture}[
  level distance = 2cm,
  every node/.style  = {circle, draw, minimum size = 14mm},
  rightChild/.style  = {circle, draw, fill = gray!30, minimum size = 15mm},
  leftlabel/.style   = {midway, xshift = -12pt, draw = none, inner sep = 1pt},
  rightlabel/.style  = {midway, xshift =  18pt, draw = none, inner sep = 1pt},
  level 1/.style     = {sibling distance = 5cm},
  level 2/.style     = {sibling distance = 3cm},
  level 3/.style     = {sibling distance = 2cm},
]
\node {$\P$}
  child { node {$\P_1$}
      edge from parent node[leftlabel] {$p_1^\star$}
  }
  child { node[rightChild] {$\bigcup\limits_{i>1} \P_i$}
    child { node {$\P_2$}
      edge from parent node[leftlabel] {$p_2^\star$}
    }
    child { node[rightChild] {$\P_3 \cup \P_4$}
      child { node {$\P_3$}               edge from parent node[leftlabel]  {$p_3^\star$} }
      child { node {$\P_4$}   edge from parent node[rightlabel] {$1-p_3^\star$} }
      edge from parent node[rightlabel] {$1-p_2^\star$}
    }
    edge from parent node[rightlabel] {$1-p_1^\star$}
  };
\end{tikzpicture}
	}
	\caption{Equivalent partition tree associated with \ac{EPSR} with $c=4$. Each node represents a partition. Empty nodes are associated with partitions $\P_i$, $i=1,\dots, c$, whereas shaded nodes are associated with unions of partitions (e.g., $ \P_3 \cup \P_4$).
    The partitioning probabilities are indicated on the edges of the tree.
	\label{fig:tree_epsr_rounds} }
\end{figure}
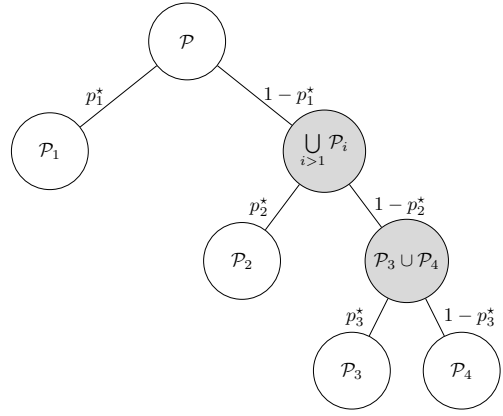

From this description, it is apparent that the number of communication rounds in \ac{EPSR} with general $c\geq2$ is equivalent to the depth of an equivalent partition tree with branching (partitioning) factor 2. However, the partition schedule in this tree is not always the same. In particular, we encounter $c-1$ different binary branchings, where the left branch has probability $p_i^\star$ and the right one $1-p_i^\star$, with  $p_i^\star = \frac{p_i}{1-\sum_{j=1}^{i-1} p_j}$, $i=1,\dots, c-1$. 
We can represent the partitioning occurring in \ac{EPSR} with arbitrary $c$ by means of an equivalent partition tree with binary branching and depth $c-1$, leading to $2^{c-1}$ leaf nodes. See Fig.~\ref{fig:tree_epsr_rounds}. In the figure, we showed partitions $\P_i$, $i=1,\dots, c$ as empty nodes, whereas nodes shaded in gray represent the union of 2 or more of these partitions.

We are interested in the expected communication round complexity, which corresponds to the expected tree depth of the equivalent partition tree (see example in Fig.\ref{fig:tree_epsr_rounds}).
Let us now introduce the notation $q_i = \max \left( p_i^\star, 1-p_i^\star\right)$, and let us define 
\[
\qmax = \max_i q_i, \text{and }
\]
\[
\qmin = \min_i q_i.
\]
Observe that the fairest branching in the equivalent partition tree will correspond to a branching with probability $\qmin$ and $1-\qmin$. Similarly, the most unbalanced (or unfair) branching will occur with probability $\qmax$ and $1-\qmax$.

Hence, the expected depth of the equivalent partition tree will be upper bounded by the expected depth  of a binary tree in which the partitioning in which the left branch has probability  $\qmax$ and the right one $1- \qmax$. 
According to Lemma~\ref{lem:psr_round}, the expected number of levels of that tree is 
$ \mathcal{O}\left( \lambda^\star \log \left( \frac{\d}{\bardelta}\right) \right)$
where $\lambda^\star = - \frac{1}{\log(\qmax)}$.
This completes the proof.

\bibliographystyle{IEEEtran}

\bibliography{IEEEabrv,references}

\end{document}